\documentclass[10pt,aps,prd,onecolumn,showpacs,superscriptaddress,longbibliography,nofootinbib]{revtex4-2}
\usepackage{mathrsfs,amsmath,amsthm,latexsym,amssymb,amsfonts,epsfig,cancel,enumerate,graphicx,txfonts,diagbox}
\usepackage[utf8]{inputenc}  
\usepackage[T1]{fontenc}     
\usepackage{lmodern}         
\usepackage{multirow}
\usepackage[T1]{fontenc}
\usepackage[utf8]{inputenc}
\usepackage[colorlinks=true,linkcolor=blue,citecolor=blue,urlcolor=blue]{hyperref}
\usepackage{orcidlink}

\usepackage{xcolor}
\usepackage{booktabs}
\usepackage{graphicx}
\usepackage{subcaption}
\definecolor{navy}{RGB}{0,0,150}
\usepackage{appendix}
\allowdisplaybreaks

\newcommand{\RGU}{Department of Physics, The Assam Royal Global University, Guwahati-781035, Assam, India}
\newcommand{\UCC}{Centro de Investigaci\'{o}n en Ciencias del Espacio y F\'{i}sica Te\'{o}rica (CICEF), Universidad Central de Chile, La Serena 1710164, Chile} 
\newcommand{\ABTU}{Department of Physics, Al-Hussein Bin Talal University, 71111, Ma’an, Jordan}

\begin{document}
\title{Observational imprints and quasi-Periodic oscillations of magnetically charged anti-de Sitter black holes}

\author{Faizuddin Ahmed\orcidlink{0000-0003-2196-9622}}
\email{faizuddinahmed15@gmail.com}
\affiliation{\RGU}

\author{Mohsen Fathi\orcidlink{0000-0002-1602-0722}}
\email{mohsen.fathi@ucentral.cl}
\affiliation{\UCC} 

\author{Ahmad Al-Badawi\orcidlink{0000-0002-3127-3453}}
\email{ahmadbadawi@ahu.edu.jo}
\affiliation{\ABTU}

\date{\today}

\begin{abstract}

In this work, we investigate the astrophysical observable signatures of a magnetically charged Anti-de Sitter (AdS) black hole in the framework of string-inspired Euler–Heisenberg theory. We focus on photon trajectories, the photon sphere, and the corresponding black hole shadow. We derive the photon sphere radius and shadow radius and show that they exhibit clear deviations from the standard Schwarzschild and Schwarzschild–AdS cases. In particular, both radii decrease monotonically as the magnetic charge parameter $Q_m$ increases, indicating that magnetic charge significantly modifies the light propagation region around the black hole.
We further analyze the dynamics of neutral and charged test particles and compute the associated epicyclic frequencies. Using the effective potential method, we obtain analytical expressions for the specific energy and specific angular momentum of particles in stable circular orbits and determine the location of the innermost stable circular orbit (ISCO). Our results show that the presence of $Q_m$ shifts the ISCO radius and modifies the orbital structure of the spacetime. We also calculate the radial, vertical, and orbital oscillation frequencies and find noticeable deviations from the classical Schwarzschild predictions.
Finally, we confront the model with observational twin–peak quasi-periodic oscillation (QPO) data from stellar–mass, intermediate–mass, and supermassive black hole candidates. A two–dimensional $\Delta\chi^2$ analysis in the $(r,Q_m)$ parameter space indicates that the best-fit configuration corresponds to $Q_m = 0$ for all considered sources, while finite values remain allowed within statistical confidence levels. At the $1\sigma$ level, we obtain an upper bound $Q_m/M \lesssim 0.2$ in the explored domain. These results suggest that although magnetic charge produces clear theoretical deviations in photon and particle dynamics, current QPO observations provide only moderate constraints on its magnitude.\\\\

{\bf Keywords:} charged black holes; null geodesics; photon sphere; black hole shadow; QPOs; fundamental frequencies 

\end{abstract}

\maketitle

\tableofcontents

\section{Introduction}

The study of black holes and their interactions with the surrounding astrophysical environment has become a cornerstone of modern theoretical astrophysics. Black holes provide a unique laboratory for exploring gravitational dynamics, high-energy processes, and the interplay between gravity and fundamental fields in the strong-field regime. Among the various black hole solutions, the Schwarzschild black hole represents the simplest and most fundamental model. It describes a static, non-rotating, and spherically symmetric spacetime solution of Einstein’s field equations in vacuum \cite{AE1916}. Because of its simplicity, the Schwarzschild geometry serves as a theoretical baseline for investigating the properties of more complex compact objects and their interactions with matter and radiation in astrophysical environments \cite{Rayibaev2020,Narzilloev2020, Rayibaev2022}.

A fundamental property of classical black holes is encapsulated in the no-hair theorem, which states that stationary black hole solutions in General Relativity (GR) are completely characterized by a small set of conserved quantities that can be measured by a distant observer. In its simplest form, the theorem implies that black holes are fully described by only three parameters: mass, angular momentum, and electromagnetic charge. Consequently, no additional independent degrees of freedom exist that could encode further information about the matter that formed the black hole. According to this conjecture, the spacetime geometry outside a black hole cannot support additional fields that produce independent physical characteristics, commonly referred to as “hair.” In this context, primary hair corresponds to new degrees of freedom that are not associated with the conserved parameters of the black hole, whereas secondary hair arises from nontrivial field configurations that are determined by the existing black hole parameters. The no-hair theorem therefore restricts the possible external field configurations around black holes and applies to a wide class of fields, including scalar, gauge, and tensor fields.

One of the most important charged black hole solutions is the Reissner-Nordström spacetime, which arises from the coupled Einstein-Maxwell equations and describes a static, spherically symmetric black hole carrying an electric or magnetic charge \cite{Grunau2011}. This solution represents a natural extension of the Schwarzschild geometry when electromagnetic fields are present. In the standard formulation, the electromagnetic field obeys Maxwell’s linear theory, where the field equations are linear in the electromagnetic field tensor. However, several theoretical developments suggest that the classical Maxwell description may not remain valid in regimes of extremely strong electromagnetic fields, such as those that may occur near compact objects. These spacetimes \cite{Grunau2011} were introduced by Plebanski et al. \cite{Plebanski1976}, and later revisited by Griffiths et al. \cite{Griffiths2006}. Theoretically, they describe stars or black holes that can additionally be characterized by acceleration, angular momentum, the NUT charge, and a cosmological constant.

To address such limitations, nonlinear electrodynamics (NED) theories have been developed, in which the electromagnetic field exhibits nonlinear self-interaction. Unlike Maxwell’s theory, where the superposition principle holds and the field equations are linear, nonlinear electrodynamics introduces higher-order corrections that modify electromagnetic dynamics in strong-field regimes. In these theories, the electromagnetic Lagrangian depends nonlinearly on the electromagnetic invariants, leading to modified field equations and potentially new black hole solutions. These nonlinear interactions may significantly influence the structure and physical properties of charged black holes.

The first well-known model of nonlinear electrodynamics is the Born–Infeld (BI) theory, proposed by Born and Infeld in 1934 \cite{Born1934}. The primary motivation for this theory was to resolve the problem of the infinite self-energy of a point-like electron that arises in classical electrodynamics. By introducing a nonlinear modification of the electromagnetic action, Born–Infeld theory imposes an upper bound on the electric field strength, thereby yielding finite electromagnetic self-energy. Later developments in quantum electrodynamics further supported the emergence of nonlinear corrections to Maxwell’s theory. In particular, Heisenberg and Euler demonstrated that vacuum polarization effects caused by virtual charged particles lead to nonlinear corrections to the electromagnetic Lagrangian at the one-loop level \cite{Heisenber1936}. These quantum corrections naturally generate nonlinear terms in the electromagnetic field, providing additional theoretical motivation for studying nonlinear electrodynamics in gravitational contexts. The coupling of nonlinear electrodynamics with GR has therefore become an important area of research, as it can lead to novel black hole solutions and modify the behavior of electromagnetic fields in strong gravitational regimes. Such models have been widely used to construct regular black hole solutions, investigate the influence of magnetic charge, and explore deviations from classical Einstein–Maxwell black hole geometries.

The study of magnetically charged black holes in nonlinear electrodynamics (NED) has attracted considerable attention in recent years. These solutions provide a natural framework for constructing regular spacetimes that avoid the central singularities typical of classical black holes, while simultaneously preserving electromagnetic duality and respecting the fundamental principles of GR \cite{Allahyari2020, Wen2023, Stuchlik2019}. Nonlinear electromagnetic theories extend Maxwell’s linear electrodynamics and have deep theoretical foundations in both classical and quantum physics. They emerge as effective descriptions in contexts such as quantum chromodynamics, string theory corrections, and various unified field theories \cite{Fradkin1985}. When coupled to Einstein’s gravity, NED models can generate regular black hole solutions in which the nonlinear electromagnetic stress-energy tensor effectively plays the role of exotic matter, smoothing out the central singularity \cite{Ayon-Beato1999, Balart2014, Bronnikov2018}. Magnetically charged configurations are particularly intriguing, as they maintain electromagnetic duality and often exhibit enhanced stability compared to electrically charged black holes, especially in astrophysical environments where electric charge is rapidly neutralized \cite{Breton2005, Gonzalez2009}. Recent developments include the introduction of two-parameter NED models capable of supporting magnetically charged black holes \cite{Kruglov2017}, as well as magnetically charged Euler–Heisenberg black holes with scalar hair, which further enrich the landscape of physically viable solutions \cite{Karakasis2022}. These studies demonstrate that magnetically charged black holes in nonlinear electrodynamics not only provide singularity-free spacetimes but also offer a versatile platform for exploring strong-field gravitational and electromagnetic phenomena in both theoretical and astrophysical settings.

Recent observations, including the first images of the supermassive black holes at the centers of the elliptical galaxy M87 and the Milky Way (Sgr~A*) \cite{EHTL1,EHTL4,EHTL6,EHTL12,EHTL16} and the detection of gravitational waves by the LIGO-Virgo collaboration \cite{LIGO1,LIGO2,LIGO3}, have provided unprecedented tests of GR in the strong-field regime. Gravitational waves emitted by binary compact objects carry detailed information about the system, with their waveform strongly influenced by the objects’ spin, mass, and possible nonlinear dynamics. These observations also open new avenues for testing modified and alternative theories of gravity, particularly when combined with X-ray observations of active galactic nuclei (AGN) \cite{Bambi2016,Zhou2018,Tripathi2019}. Additionally, high-precision Very Long Baseline Interferometry (VLBI) measurements, such as those performed by the GRAVITY instrument, have tracked the highly relativistic motion of matter and the S2 star near Sgr~A*, providing further stringent tests of GR in the vicinity of supermassive black holes. These multi-messenger observations collectively offer a unique window into strong gravitational fields, enabling constraints on both classical and quantum-inspired models of black holes.

In classical GR, isolated black holes do not emit electromagnetic radiation, and therefore cannot be observed directly. However, in binary systems, the gravitational field of a black hole strongly influences the surrounding accretion disk, leading to observable electromagnetic signals. One important phenomenon in this context is the appearance of QPOs, detected through Fourier analysis of X-ray flux from accretion disks in (micro)quasars \cite{ Ingram2016, Bambi2017b, StuchlikKolos2015}. QPOs are usually classified as high-frequency (HF, $\sim$0.1--1~kHz) or low-frequency (LF, $<0.1$~kHz), and have been observed around black holes, neutron stars, and white dwarfs in binary systems. HF QPOs in low-mass X-ray binaries (LMXBs) exhibit diverse behaviors, suggesting different underlying physical mechanisms, while in neutron stars they are often linked to magnetic field strength and rotation \cite{Torok2019}. In black hole systems, QPOs are commonly interpreted as arising from (quasi-)harmonic oscillations of matter in stable circular orbits, where the emitted radiation reflects the fundamental orbital and epicyclic frequencies. Although numerous QPOs have been detected with high precision, their exact physical origin remains an open problem. Nevertheless, they provide a powerful tool for probing the innermost regions of accretion disks and for testing gravity in the strong-field regime \cite{Stuchlik2011b,Torok2011b,JRayimbaev2021}.

Microquasars are binary systems composed of a compact object—either a black hole or a neutron star—and a companion main-sequence star, and are observed as X-ray binaries. These systems provide natural laboratories for testing gravitational theories under extreme conditions. QPOs were first detected through power spectral analysis of X-ray flux from a low-mass X-ray binary (LMXB) \cite{Angelini1989}, and since then they have become an important probe of the gravitational properties of compact objects \cite{Torok2007}. In particular, twin-peaked QPOs exhibiting a characteristic $3\!:\!2$ frequency ratio have been observed in several black hole candidates \cite{Torok2007,TorokStuchlik2005,StuchlikKolos2016}. Such frequency pairs are especially useful for constraining models of accretion disk oscillations and the underlying spacetime geometry.

In light of these considerations, in the present work we study the observational imprints and QPO properties of a magnetically charged AdS black hole in string-inspired Euler–Heisenberg theory \cite{AB2024}. We analyze null geodesics, the photon sphere, the shadow radius, and the dynamics of neutral and charged particles in circular motion. Furthermore, motivated by the direct relation between epicyclic frequencies and QPO models, we confront the theoretical predictions with observed twin–peak QPO data from stellar–mass, intermediate–mass, and supermassive black hole candidates. By performing a statistical fitting procedure in the $(r,Q_m)$ parameter space, we investigate whether the magnetic charge parameter $Q_m$ leaves observable signatures in timing data and determine the corresponding confidence regions.

The paper is organized as follows: In Section \ref{sec:2}, we introduce the magnetically charged AdS black hole solution. In Section \ref{sec:3}, we present a detailed study of geodesic motion, analyzing null trajectories, effective potentials, and the black hole shadow. In Section \ref{sec:4}, we investigate the dynamics of neutral test particles using the Hamiltonian formalism. In Section \ref{sec:5}, we study the motion of charged particles and the conditions for circular orbits. In Section \ref{sec:6}, we analyze the oscillations of such particles and compute the relevant frequencies for QPO models in Section \ref{sec:QPOs}. Finally, in Section \ref{sec:7}, we summarize our findings.

\section{Theoretical framework of AdS-Black Hole } \label{sec:2}

In the geometrized unit system $(c = G = 1)$, the Einstein-Hilbert action reads \cite{AB2024}
\begin{equation}
S = \frac{1}{16\pi} \int d^4x \sqrt{-g} \Big[\mathcal{R}-2 \nabla^\mu \phi \nabla_\mu \phi
- e^{-2\phi} \mathcal{F}^2- f(\phi)\,\left(2\alpha F^\alpha{}_\beta F^\beta{}_\gamma F^\gamma{}_\delta F^\delta{}_\alpha- \beta F^4\right)-\mathcal{D}(\phi)\Big].\label{aa1}
\end{equation}
Such a field theoretic gravitational action admits the Gibbons-Maeda-Garfinkle-Horowitz-Strominger (GMGHS) A(dS) black hole \cite{DG1992,GWG1998} as an exact solution when $f(\phi)=0$. In~(\ref{aa1}), $\mathcal{R}$ is the Ricci scalar, $\mathcal{D}(\phi)$ is a dilaton scalar potential, $\mathcal{F}^2 \equiv F_{\mu\nu}\,F^{\mu\nu} \sim \bm{E}^2 - \bm{B}^2$ is the usual Faraday scalar, and $F^4 \equiv F_{\mu\nu}\,F^{\mu\nu}\,F_{\alpha\beta}\,F^{\alpha\beta}$, where $F_{\mu\nu}$ stands for the usual field strength $F_{\mu\nu} = \partial_\mu A_\nu - \partial_\nu A_\mu$ and $\alpha,\,\beta$ are coupling constants of the theory, with dimensions (length)$^2$. The scalar field $\phi$ and the associated scalar function $f(\phi)$ are both dimensionless.

In this work, we consider the coupling function $f(\phi)$ and the scalar potential $\mathcal{D}(\phi)$ given by \cite{AB2024}
\begin{align}
    &f(\phi)=-\frac{1}{2}\left(e^{-2\phi}+e^{2\phi}+4\right) \equiv -\left[3 \cosh (2\phi)+2\right],\label{aa2}\\
    &\mathcal{D}(\phi)=\frac{2\Lambda}{3} \left[\cosh (2\phi)+2\right].\label{aa3}
\end{align}

Thereby, the metric corresponding to the magnetically charged A(dS) black hole is described by the following line element \cite{AB2024}
\begin{equation}
    ds^2 = -f(r)dt^2 + \dfrac{dr^2}{f(r)} + D^2(r) (d\theta ^2 + \sin ^2{\theta }\,d\varphi ^2),\label{metric}
\end{equation}
where the metric functions function are given by
\begin{align}
     f(r) = 1-\frac{2 M}{r}-\frac{2 \epsilon Q^4_m}{r^3 (r-Q^2_m/M)^3}-\frac{\Lambda}{3}\,r \left(r-\frac{Q^2_m}{M}\right), \qquad D^2(r)=r\,\left(r-\frac{Q^2_m}{M}\right),
    \label{function}
\end{align}
corresponding to the choice
\begin{equation}
\phi(r)=-\frac{1}{2}\ln\!\left(1-\frac{Q^2_m}{M r}\right),\qquad A_{\mu}=(0,\,0,\,0,\,Q_m \cos \theta).
    \label{eq:phi,A}
\end{equation}
Here $Q_m$ is the magnetic charge, $M$ represent mass of the black hole. This solution (\ref{metric}) describes a black hole with a single horizon when $\epsilon=1$, whereas  for $\epsilon=-1$,  the  black  hole  horizons can  range from  two  to  none.  This  indicates  that  the  magnetically charged black hole possesses different horizon structures. It is worth noting that the radial coordinate lies in the interval $Q^2_m/M \leq r < \infty$ such that the metric component $D(r) \in [0, \infty)$. In the limit $\epsilon=0$, the metric (\ref{metric}) reduces to the Garfinkle-Horowitz-Strominger (GHS) AdS black hole solution \cite{DG1992}. Furthermore, in the limit $Q_m=0$, corresponding to the absence of magnetic charge, we recover Schwarzschild AdS black hole.

\begin{figure}[ht!]
    \centering
    \includegraphics[width=0.45\linewidth]{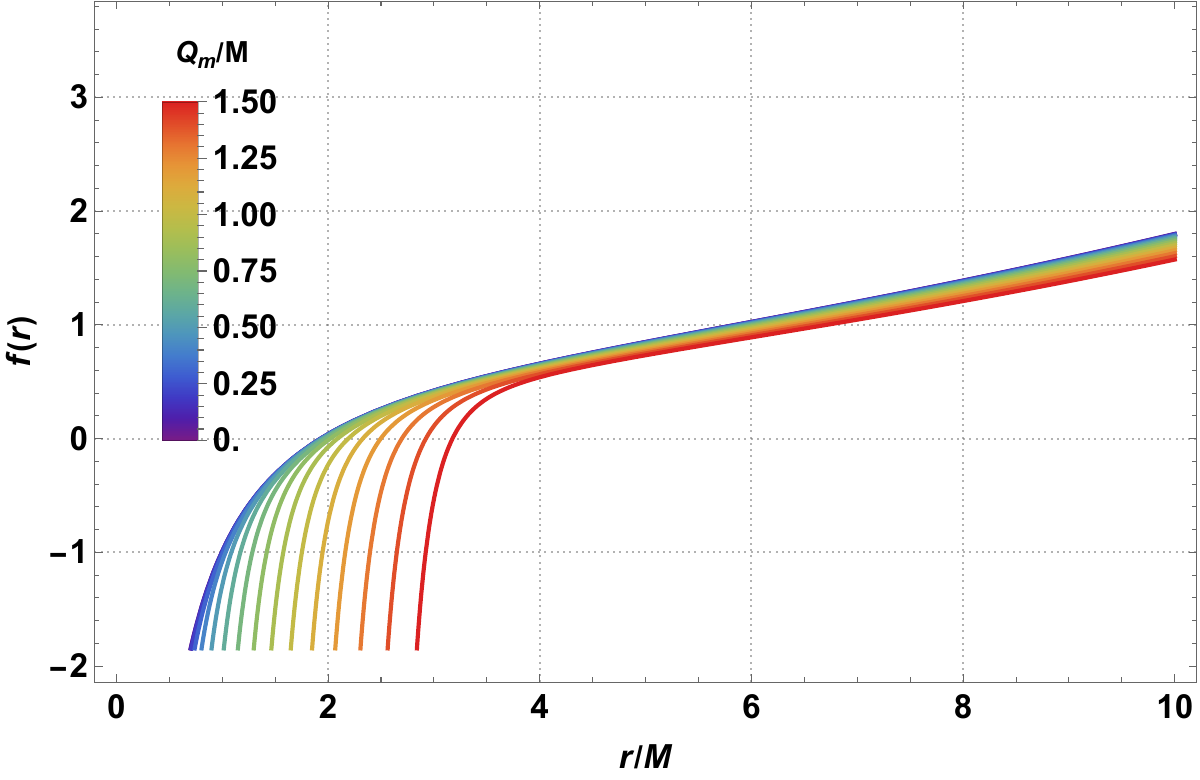}\qquad
    \includegraphics[width=0.45\linewidth]{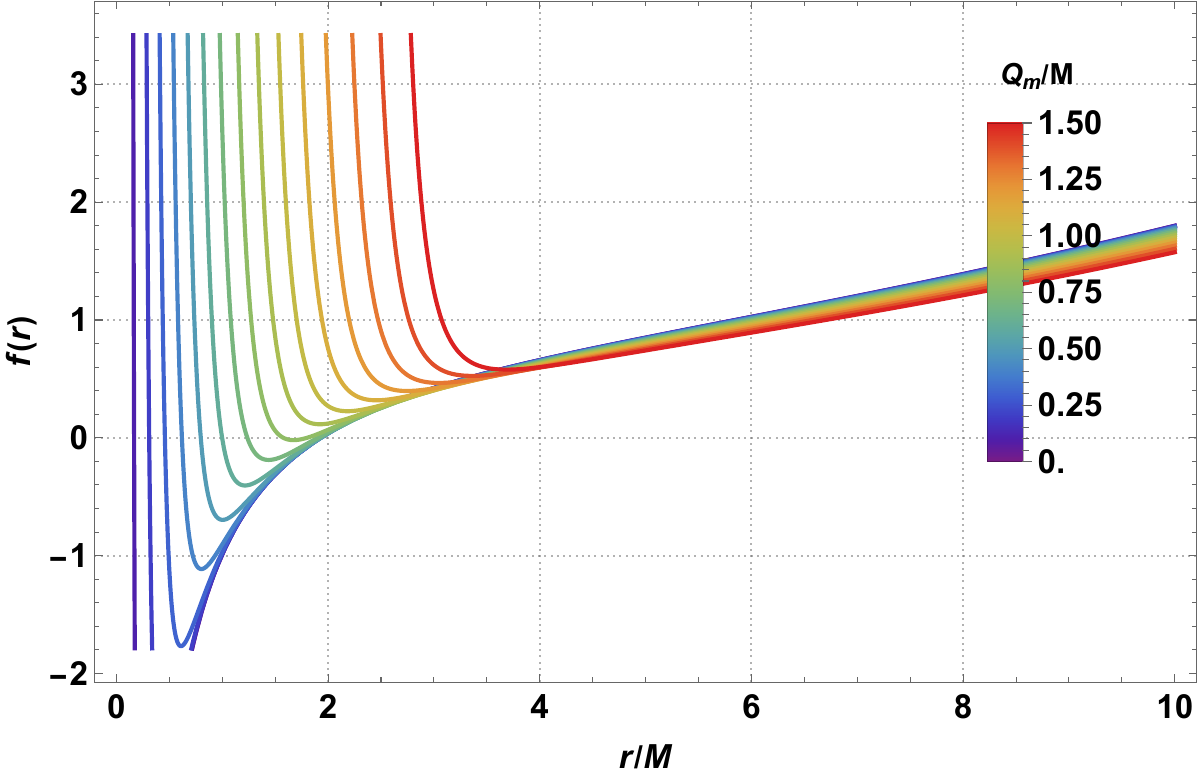}\\
    (i) $\epsilon=1$ \hspace{8cm} (ii) $\epsilon=-1$
    \caption{ Behavior of the metric function $f(r)$ Eq. (\ref{function}) for a magnetically charged AdS black holes. Here $M=1,\,\,\Lambda=-0.02$.}
    \label{fig:metric}
\end{figure}

Noted that at radial large distance, the metric components behaves as
\begin{equation}
    \lim_{r \to \infty} f(r) \simeq 1-\frac{\Lambda}{3}r^2,\qquad \lim_{r \to \infty} \frac{D^2(r)}{r^2} =1.\label{case}
\end{equation}
This indicates that the magnetically charged black hole metric is asymptotically AdS background. For zero cosmological constant, the space-time is asymptotically flat space at spatial infinity. \\ Figure  \ref{fig:metric} shows the radial profile of the metric
function $f(r)$ for $\epsilon =\pm\,1$  as $Q$ increases from 0 to 1. The 
left panel illustrates the  case $\epsilon =1$, 
the solution (\ref{metric}) describes
a black hole with a single horizon. In the right panel $\epsilon =-1$,  the solution horizons can range
from two to none.

\section{Photon Sphere and Shadow}\label{sec:3}

Null geodesics play a fundamental role in black hole physics, as they govern the motion of photons in curved spacetime. Through the analysis of null geodesics, one can determine key observational and geometrical features of a black hole, such as the radius of the photon sphere, the structure and size of the black hole shadow, photon trajectories, and the effective forces experienced by photons in a given gravitational field \cite{Chandrasekhar1984, Wald1984}. These trajectories encode crucial information about the underlying spacetime geometry and therefore serve as powerful probes of strong gravitational fields.

Recent high-resolution observations by the Event Horizon Telescope (EHT) collaboration \cite{EHTL1,EHTL4,EHTL6,EHTL12,EHTL16,EHTL17} have significantly deepened our understanding of black holes and have opened new avenues for testing gravitational theories in the strong-field regime. The first horizon-scale images of supermassive black holes have provided direct observational evidence of shadow structures, which are intimately connected to the behavior of null geodesics near the event horizon. These groundbreaking results have further stimulated theoretical investigations into photon motion and black hole optics.

The photon sphere is a spherical region surrounding a black hole where gravity is strong enough to force photons to travel in unstable circular orbits. It represents the boundary between photons that escape to infinity and those that are captured by the black hole. The radius of the photon sphere depends on the spacetime geometry and is therefore sensitive to parameters such as mass, charge, spin, and possible modifications to GR. Since the apparent shape and size of the black hole shadow are directly related to the photon sphere, studying null geodesics provides essential insights into observable signatures of compact objects and offers a direct link between theory and astrophysical observations.

The space-time (\ref{metric}) can be expressed as $ds^2=g_{\mu\nu} dx^{\mu} dx^{\nu}$, where the metric tensor $g_{\mu\nu}$ with $\mu,\nu=0,...,3$ is given by
\begin{equation}
    g_{\mu\nu}=\mbox{diag}\left(-f(r),\,\frac{1}{f(r)},\,D^2(r),\,D^2(r) \sin^2 \theta\right),\label{bb1} 
\end{equation}

The Lagrangian density in terms of the metric tensor $g_{\mu\nu}$ is given by
\begin{equation}
    \mathbb{L}=\frac{1}{2}\,g_{\mu\nu}\, \dot x^{\mu}\,\dot x^{\nu},\label{bb2}
\end{equation}
where dot represents ordinary derivative w. r. to an affine parameter $\lambda$ along geodesics.

Considering the geodesic motion in the equatorial plane defined by $\theta=\pi/2$, the Lagrangian density function (\ref{bb2}) using (\ref{bb1}) explicitly written as
\begin{equation}
    \mathbb{L}=\frac{1}{2}\left[-f(r)\,\dot t^2+\frac{1}{f(r)} \dot r^2+D^2(r)\,\dot \phi^2 \right].\label{bb3}
\end{equation}

The Lagrangian density function is independent of temporal coordinate $t$ and angular coordinate $\theta$. Therefore, there exist two conserved quantities associated with these coordinates. These are the conserved energy and conserved angular momentum and are given by \cite{Al-Badawi2023,ALBADAWI2025185,FA7,FA11}
\begin{equation}
    \mathrm{E}=f(r)\,\dot t,\label{bb4}
\end{equation}
And
\begin{equation}
    \mathrm{L}=D^2(r)\,\dot \phi.\label{bb5}
\end{equation}

Based on the above, the equation of motion for photon trajectories for the radial coordinate can be expressed as
\begin{equation}
    \dot r^2+V_{\rm eff}=\mathrm{E}^2,\label{bb6}
\end{equation}
where $V_{\rm eff}$ is the effective potential governing the photon dynamics and is given by
\begin{equation}
    V_{\rm eff}=\frac{\mathrm{L}^2}{D^2(r)}\,f(r)=\frac{\displaystyle 1-\frac{2 M}{r}-\frac{2 \epsilon Q^4_m}{r^3 (r-Q^2_m/M)^3}-\frac{\Lambda}{3}\,r \left(r-\frac{Q^2_m}{M}\right)}{\displaystyle r (r-Q^2_m/M)}\,\mathrm{L}^2.\label{bb7}
\end{equation}

\begin{figure}
    \centering
    \includegraphics[width=0.45\linewidth]{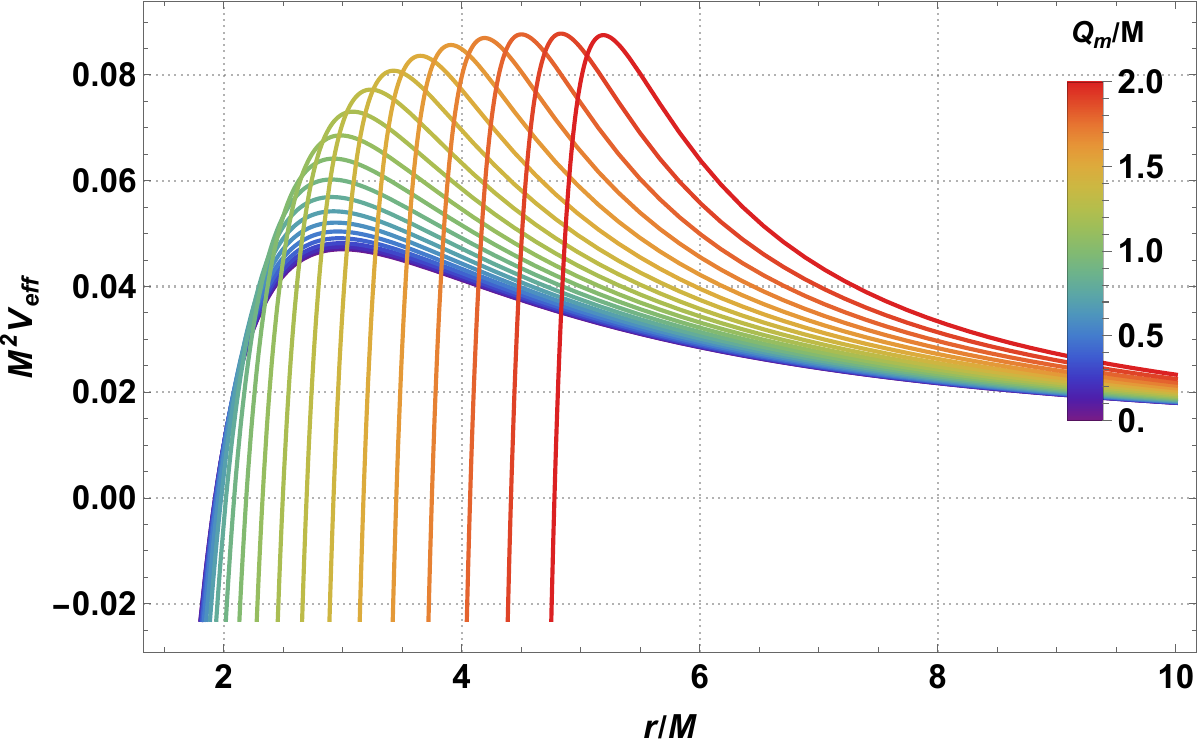}
    \includegraphics[width=0.45\linewidth]{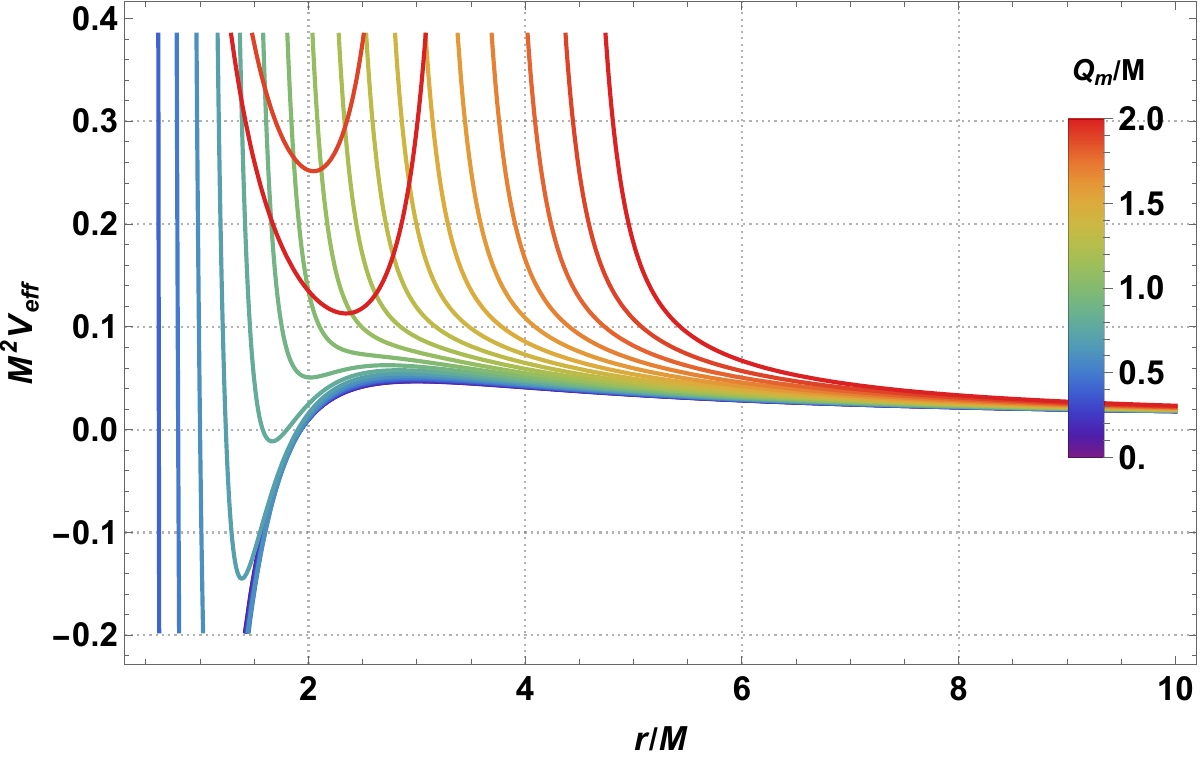}\\
    (i) $\epsilon=1$ \hspace{6cm} (ii) $\epsilon=-1$
    \caption{Behavior of the squared mass times the effective potential as a function of the dimensionless radial coordinate $r/M$, for different values of the magnetic charge parameter $Q_m/M$.}
    \label{fig:potential}
\end{figure}

From the above potential expression, we observe that the effective potential governing the photon dynamics is influenced by the geometric parameters that alter the space-time curvature. These include the the magnetic charge $Q_m$, the black hole mass $M$, the cosmological constant $\Lambda$. Moreover, the conserved angular momentum $\mathrm{L}$ alter this effective potential. In Figure \ref{fig:potential}, we illustrate the behavior of the effective potential by varying the magnetic charge $Q_m$.

Next, we study the circular orbits in  which the photon particles revolving around the black hole. For circular null orbits, the conditions $\frac{dr}{d\lambda}=0$ and $\frac{d^2r}{d\lambda^2}=0$ must be satisfied. From Eq. (\ref{bb6}), we find the following relations:
\begin{equation}
    \mathrm{E}^2=V_{\rm eff}=\frac{\mathrm{L}^2}{D^2(r)}\,f(r)\Longrightarrow \beta_c=\frac{D(r)}{\sqrt{f(r)}}\Big{|}_{r=\text{const.}},\label{bb8}
\end{equation}
where $\beta_c$ is the critical impact parameter for photon particles at fixed radii for circular orbits. And the second condition is
\begin{equation}
    \frac{d}{dr}\left(\frac{f(r)}{D^2(r)}\right)=0.\label{bb9}
\end{equation}

The relation (\ref{bb9}) gives us the photon sphere radius $r=r_s$ satisfying the following polynomial relation as,
\begin{align}
3M+\frac{Q_m^2}{2M}-\frac{2Q_m^2}{r}-r+\frac{\displaystyle 4 Q_m^4\left(2r-\frac{Q_m^2}{M}\right)\epsilon}
{\displaystyle r^3\left(r-\frac{Q_m^2}{M}\right)^3}= 0.\label{bb10}
\end{align}

The exact analytical solution of the above polynomial yields the photon sphere radius $r_s$. Noted that the exact analytical solution is quite a challenging task due to the higher order polynomial (7th order). However, one can determine the numerical results by selecting suitable values of the magnetic charge $Q_m$. In the limit $Q_m=0$, one finds $r_s=3M$, the result for the standard Schwarzschild black hole case. In Table \ref{tab:1}, we presented numerical results for the photon sphere radius by varying $Q_m$.

\begin{table}[ht!]
\centering
\begin{tabular}{|c|c|c|}
\hline
$Q_m/M$ & $r_{\mathrm{ph}}/M\quad (\epsilon = 1)$ & $r_{\mathrm{ph}}/M\quad (\epsilon = -1)$ \\
\hline
0.1 & 2.998333 & 2.998326 \\
0.2 & 2.993329 & 2.993218 \\
0.3 & 2.984994 & 2.984391 \\
0.4 & 2.973391 & 2.971287 \\
0.5 & 2.958763 & 2.952894 \\
0.6 & 2.941786 & 2.927332 \\
0.7 & 2.924128 & 2.890789 \\
0.8 & 2.909538 & 2.834094 \\
0.9 & 2.905595 & 2.724682 \\
1.0 & 2.924915 & 2.551854 \\
1.1 & 2.982143 & 2.815914 \\
1.2 & 3.086158 & 3.091723 \\
1.3 & 3.236461 & 3.380350 \\
1.4 & 3.427456 & 3.682770 \\
1.5 & 3.653251 & 3.999860 \\
\hline
\end{tabular}
\caption{Numerical values of the photon sphere radius $r_{\mathrm{ph}}/M$ for different magnetic charge values $Q_m/M$ and for $\epsilon=\pm1$, rounded to six decimal places.}
\label{tab:1}
\end{table}

Next, we determine the size of the shadow cast by the black hole, showing how the magnetic charge alter this. For a static observer located at position $r_O$, the angular size of the shadow is given by \cite{Volker2022} 
\begin{equation}
   \sin \theta_{\rm sh}=\frac{h(r_s)}{h(r_O)},\qquad h(r)=\frac{D(r)}{\sqrt{f(r)}}.\label{bb12}
\end{equation}
The shadow radius is therefore given by
\begin{align}
    R_{\rm sh} \simeq r_O\,\theta_{\rm sh}&=r_O\,\frac{D(r_s)}{D(r_O)}\,\sqrt{\frac{f(r_O)}{f(r_s)}}\nonumber\\
    &=\sqrt{\frac{r_s\,\left(r_s-Q^2_m/2M\right)}{1-Q^2_m/(2 M r_O)}}\,\sqrt{ \frac{\displaystyle 1-\frac{2 M}{r_O}-\frac{2 \epsilon Q^4_m}{r^3_O (r_O-Q^2_m/M)^3}-\frac{\Lambda}{3}\,r_O \left(r_O-\frac{Q^2_m}{M}\right)}{\displaystyle 1-\frac{2 M}{r_s}-\frac{2 \epsilon Q^4_m}{r^3_s (r_s-Q^2_m/M)^3}-\frac{\Lambda}{3}\,r_s \left(r-\frac{Q^2_m}{M}\right)}},\label{shadow}
\end{align}
where the photon sphere radius $r_s$ can be determined using the relation (\ref{bb10}).

From the above expression, we observe that the shadow radius depends on several geometric parameters including the magnetic charge $Q_m$, the black hole mass $M$ and the cosmological constant $\Lambda$ for a particular position of the static observer.

\begin{table}[ht!]
\centering
\begin{tabular}{|c|c|c|}
\hline
$Q_m/M$ & $R_{\rm sh}/M$ $(\epsilon = 1)$ & $R_{\rm sh}/M$ $(\epsilon = -1)$ \\
\hline
0.1 & 9.5297617168 & 9.5297539853 \\
0.2 & 9.5065447563 & 9.5064166955 \\
0.3 & 9.4678967139 & 9.4672093512 \\
0.4 & 9.4139630842 & 9.4116011922 \\
0.5 & 9.3451319357 & 9.3386914003 \\
0.6 & 9.2623122354 & 9.2469507923 \\
0.7 & 9.1674606312 & 9.1336433946 \\
0.8 & 9.0645110669 & 8.9934011877 \\
0.9 & 8.9607565106 & 8.8147225614 \\
1.0 & 8.8681558848 & 8.4527624706 \\
1.1 & 8.8030749572 & 8.1073556525 \\
1.2 & 8.7830628608 & 8.0459635604 \\
1.3 & 8.8216528729 & 8.1348051742 \\
1.4 & 8.9251211214 & 8.3177817130 \\
1.5 & 9.0934246917 & 8.5672940930 \\
\hline
\end{tabular}
\caption{Shadow radius for various magnetic charges $Q_m$ and $\epsilon = \pm 1$.Here $r_O/M=50,\,\Lambda=-0.003/M^2$.}
\end{table}

However, for a distant observer, $r_O \to \infty$, the shadow radius for RN-like black hole (setting the cosmological constant to be zero) reduces to
\begin{equation}
    R_{\rm sh}=\sqrt{\frac{r_s\,\left(r_s-Q^2_m/2M\right)}{\displaystyle 1-\frac{2 M}{r_s}-\frac{2 \epsilon Q^4_m}{r^3_s (r_s-Q^2_m/M)^3}}},\label{shadow2}
\end{equation}

Below, we discuss some special cases of the shadow radius given in Eq.~(\ref{shadow}).
\begin{itemize}
    \item When $Q_m = 0$, corresponding to the absence of magnetic charge of the black hole, and $\Lambda=0$, the photon sphere radius becomes $r_s=3M$. Consequently, the shadow radius simplifies as
\begin{equation}
    R^{\rm Sch}_{\rm sh}=3\sqrt{3} M\label{bb13}
\end{equation}
which is similar to the Schwarzschild black hole case.

\item When $\epsilon=0$ and $\Lambda=0$, the photon sphere radius becomes 
\begin{equation}
    r_s=\frac{1}{2}\left[3 M+\frac{Q^2_m}{2M}+\sqrt{9 M^2+\frac{Q^4_m}{4 M^2}-5 Q^2_m}\right].\label{bb14}
\end{equation}
Consequently, the shadow radius simplifies as,
\begin{equation}
    R_{\rm sh}=\sqrt{\frac{r_s\,\left(r_s-Q^2_m/2M\right)}{1-Q^2_m/(2 M r_O)}}\,\sqrt{\frac{1-\frac{2 M}{r_O}}{1-\frac{2 M}{r_s}}},\label{shadow3}
\end{equation}

For a static distant observer ($r_O \to \infty$), the shadow radius reduces to
\begin{align}
    R_{\rm sh}&=r_s\,\sqrt{\frac{r_s-Q^2_m/2M}{r_s-2 M}}\nonumber\\
    &=\frac{1}{2}\left[3 M+\frac{Q^2_m}{2M}+\sqrt{9 M^2+\frac{Q^4_m}{4 M^2}-5 Q^2_m}\right]\,\sqrt{\frac{3 M-\frac{Q^2_m}{2M}+\sqrt{9 M^2+\frac{Q^4_m}{4 M^2}-5 Q^2_m}}{- M+\frac{Q^2_m}{2M}+\sqrt{9 M^2+\frac{Q^4_m}{4 M^2}-5 Q^2_m}}}.\label{shadow4}
\end{align}
\end{itemize}

\begin{figure}[ht!]
    \centering
    \includegraphics[width=0.45\linewidth]{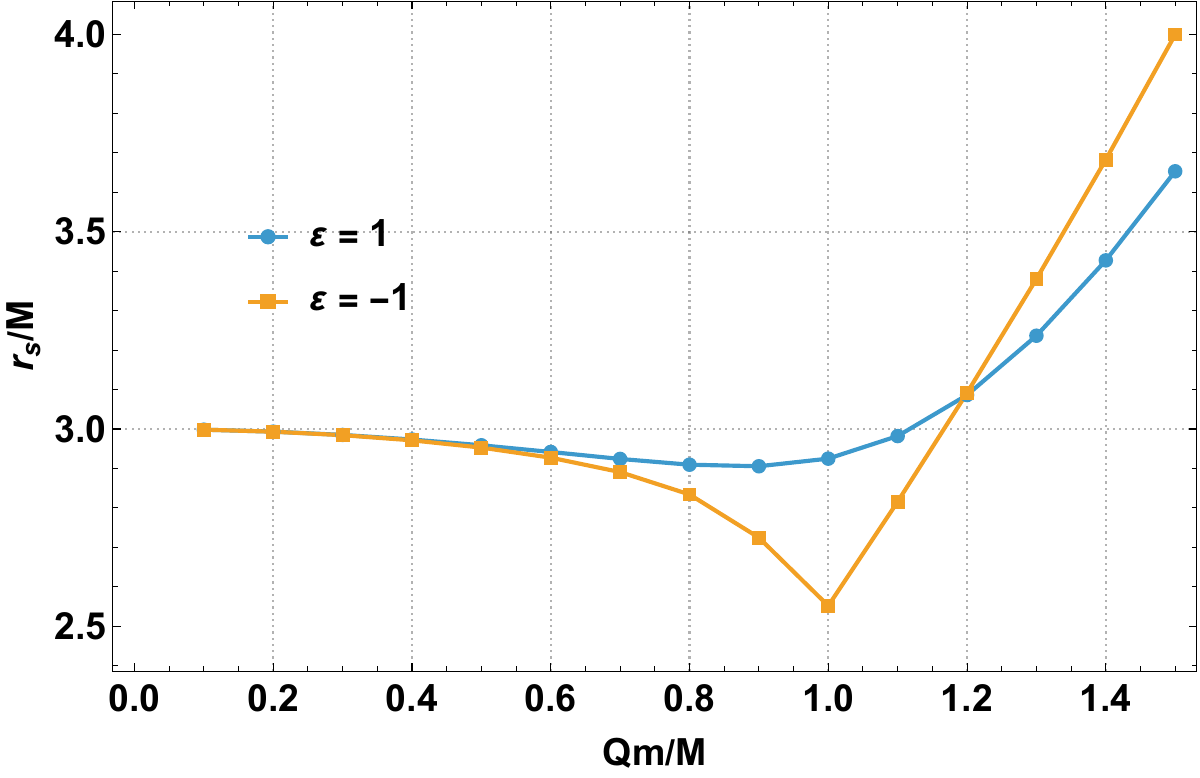}\qquad
    \includegraphics[width=0.45\linewidth]{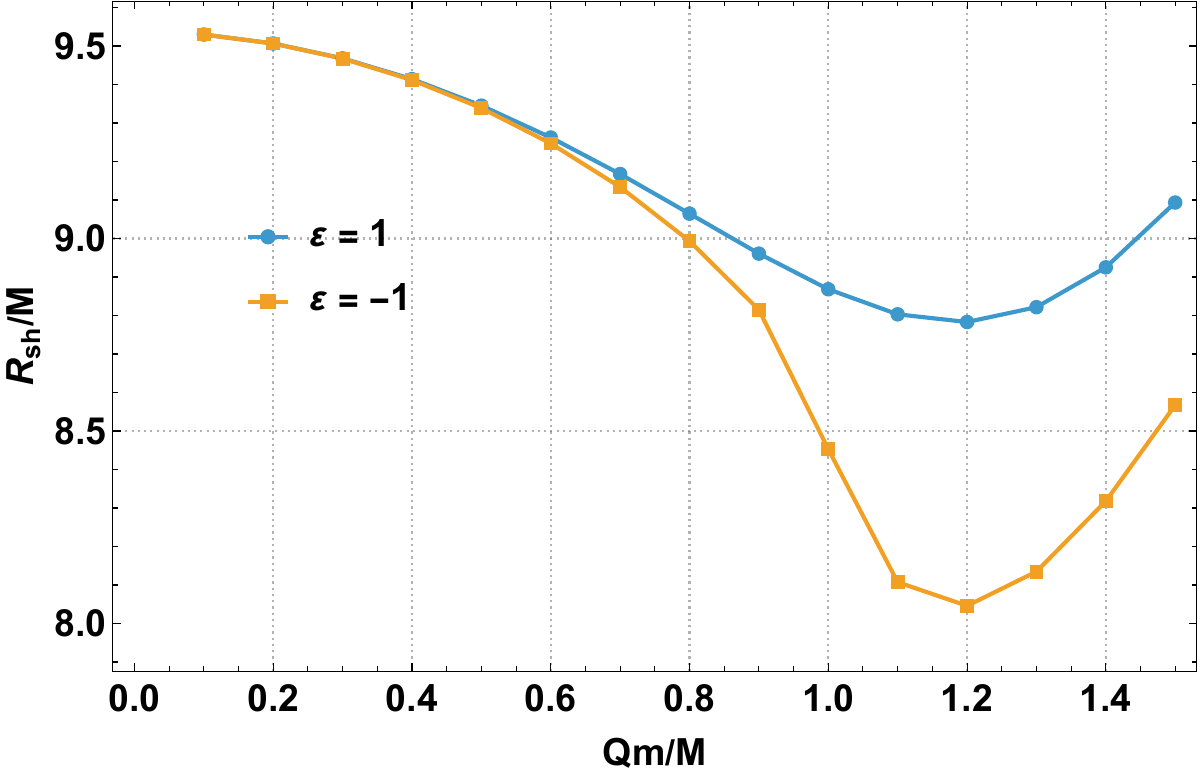}
    \caption{Photon sphere radius $r_s/M$ and shadow radius $R_{sh}/M$ by varying the magnetic charge parameter $Q_m/M$.}
    \label{fig:photon-sphere}
\end{figure}

Figure \ref{fig:photon-sphere} shows how the photon sphere radius and the shadow radius change with the magnetic charge parameter $Q_m$. As $Q_m$ increases from 0 to 1, we observe a monotonic decrease in both radii, indicating that magnetic charge has a significant effect on light trapping regions around the black hole. Overall, both branches ($\epsilon =\pm 1$) demonstrate  slightly different shadow sizes. 

\section{Dynamics of Neutral Test Particles}\label{sec:4}

In this section, we study the dynamics of neutral test particles around the selected black hole solution using the Hamiltonian formalism. We derive the effective potential that governs particle motion and analyze the conditions for circular orbits. For these orbits, we determine the specific energy and specific angular momentum, demonstrating how the metric parameters-such as the magnetic charge, black hole mass, and AdS radius-affect them. Furthermore, we examine the properties of the ISCO.

The motion of a neutral particle with mass $m$ is governed by the Hamiltonian:
\begin{equation}
    H = \frac{1}{2}g^{\mu\nu}p_{\mu}p_{\nu} + \frac{1}{2}m^2,\label{cc1}
\end{equation}
where $p^{\mu} = mu^{\mu}$ is the 4-momentum, $u^{\mu} = dx^{\mu}/d\tau$ is the 4-velocity, and $\tau$ is the proper time. The Hamiltonian equations of motion are
\begin{equation}
\frac{dx^\mu}{d\zeta} \equiv mu^\mu = \frac{\partial H}{\partial p_\mu}, \qquad 
\frac{dp_\mu}{d\zeta} = -\frac{\partial H}{\partial x^\mu},\label{cc2}
\end{equation}
where $\zeta = \tau/m$ is the affine parameter.

The static, spherically symmetric spacetime admits two Killing vectors associated with time translation and axial rotation, yielding two conserved quantities:
\begin{align}
&\frac{p_t}{m} = -f(r)\frac{dt}{d\tau} = -\mathcal{E},\label{cc3}\\
&\frac{p_\phi}{m} = D^2(r)\,\sin^2\theta\,\frac{d\phi}{d\tau} = \mathcal{L},\label{cc4}
\end{align}
where $\mathcal{E}$ and $\mathcal{L}$ are the specific energy and specific angular momentum (per unit mass), respectively.

The 4-velocity components are:
\begin{align}
    &\frac{dt}{d\tau} = \frac{\mathcal{E}}{f(r)},\label{cc5}\\
    &\frac{d\phi}{d\tau} = \frac{\mathcal{L}}{D^2(r)\,\sin^2\theta},\label{cc6}\\
    &\frac{d\theta}{d\tau} = \frac{p_{\theta}}{m\,D^2(r)},\label{cc7}\\
    &\frac{dr}{d\tau} = \sqrt{\mathcal{E}^2 - \left(\epsilon + \frac{\mathcal{L}^2}{D^2(r)\sin^2\theta} + \frac{p^2_{\theta}}{m\,D^2(r)}\right)f(r)},\label{cc8}
\end{align}
where $\epsilon = 1$ for time-like particles and $\epsilon = 0$ for light-like particles.

In our work, we focus only on time-like
particles. Therefore, for time-like particles, the Hamiltonian in Eq.~\eqref{cc1} can be re-written as:
\begin{equation}
    H = \frac{f(r)}{2}p^2_r + \frac{p^2_{\theta}}{D^2(r)} + \frac{1}{2}\frac{m^2}{f(r)}\left[U_{\rm eff}(r,\theta) - \mathcal{E}^2\right],\label{cc9}
\end{equation}
where the effective potential $U_{\rm eff}(r,\theta)$ takes the form:
\begin{align}
    U_{\rm eff}(r,\theta) &= \left(1 + \frac{\mathcal{L}^2}{D^2(r)\,\sin^2\theta}\right)\,f(r)\nonumber\\
    &=\left(1 + \frac{\mathcal{L}^2\,\left(1-\frac{Q^2_m}{M r}\right)^{-1}}{r^2\,\sin^2\theta}\right)\left[1-\frac{2 M}{r}-\frac{2 \epsilon Q^4_m}{r^3 \left(r-\frac{Q^2_m}{M}\right)^3}-\frac{\Lambda}{3}\,r \left(r-\frac{Q^2_m}{M}\right)\right].\label{cc10}
\end{align}

In the equatorial plane ($\theta = \pi/2$), the effective potential becomes:
\begin{equation}
    U_{\rm eff}(r) =\left(1 + \frac{\mathcal{L}^2\,\left(1-\frac{Q^2_m}{M r}\right)^{-1}}{r^2}\right)\left[1-\frac{2 M}{r}-\frac{2 \epsilon Q^4_m}{r^3 \left(r-\frac{Q^2_m}{M}\right)^3}-\frac{\Lambda}{3}\,r \left(r-\frac{Q^2_m}{M}\right)\right].\label{cc11}
\end{equation}
From the above potential expression , it is clear that the effective potential governing the particle dynamics is influenced by the magnetic charge $Q_m$, the black hole mass $M$, and the cosmological constant $\Lambda$. Belowe we discuss the circular orbits conditns and analyze the outcome.

\begin{figure}[ht!]
    \centering
    \includegraphics[width=0.45\linewidth]{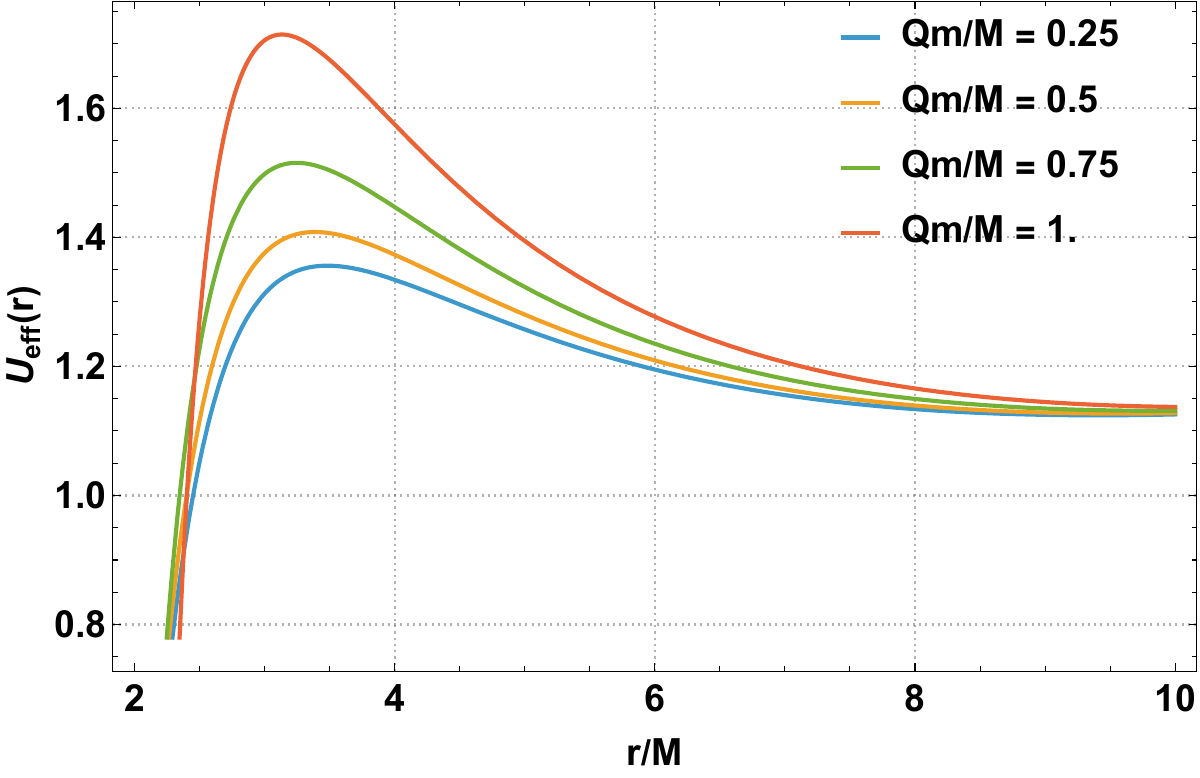}\qquad
    \includegraphics[width=0.45\linewidth]{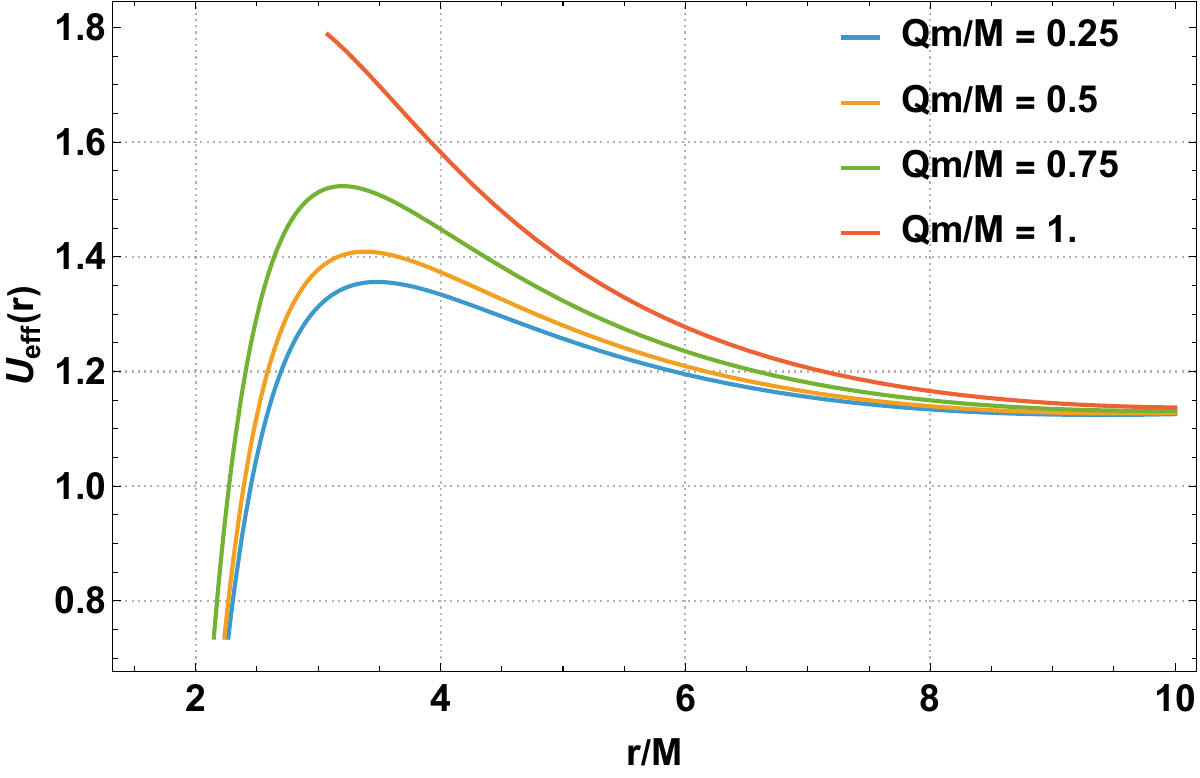}\\
    (i) $\epsilon=1$ \hspace{6cm} (ii) $\epsilon=-1$
    \caption{The behavior of the effective potential governing the particle dynamics as a function of dimensionless radial distance $r/M$ by varying the magnetic charge $Q_m/M$. Here $\Lambda=-0.0003/M^2,\,\mathcal{L}/M=5$.}
    \label{fig:placeholder}
\end{figure}

For circular motion of test particles on the equatorial plane, the following conditions must be satisfied
\begin{align}
    \mathcal{E}^2&=U_{\rm eff}(r),\label{cc12}\\
    \partial_r U_{\rm eff}&=0.\label{cc13}
\end{align}
Using potential in (\ref{cc11}) in to the above conditions and after simplification, we find the specific angular momentum $\mathcal{L}_{\rm sp}$ and the specific energy $\mathcal{E}_{\rm sp}$ of test particles as follows:
\begin{align}
    \mathcal{L}_{\rm sp}&=\sqrt{\frac{A^2(r)\,f'(r)}{A'(r)\,f(r)-A(r)\,f'(r)}}\Big{|}_{r=\mbox{const.}},\qquad \mathcal{E}_{\rm sp}=\sqrt{\frac{A'(r)\,f^2(r)}{A'(r)\,f(r)-A(r)\,f'(r)}}\Big{|}_{r=\mbox{const.}},\nonumber\\
    A(r)&=r\,(r-Q^2_m/M).\label{cc14}
\end{align}
For fixed radii $r=\mbox{const.}$ of the circular orbits, these physical quantities are influenced by the magnetic charge, $Q_m$, the black hole mass $M$, and the cosmological constant $\Lambda$.

\begin{figure}[ht!]
    \centering
    \includegraphics[width=0.45\linewidth]{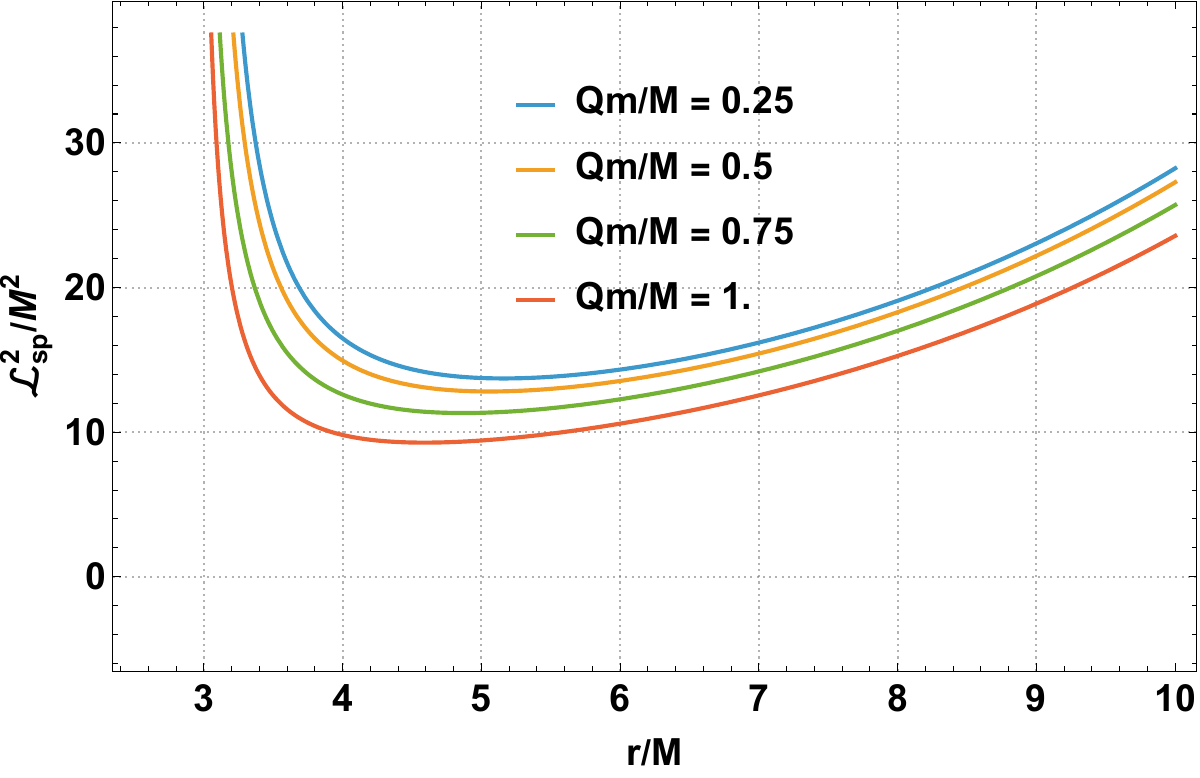}\qquad
    \includegraphics[width=0.45\linewidth]{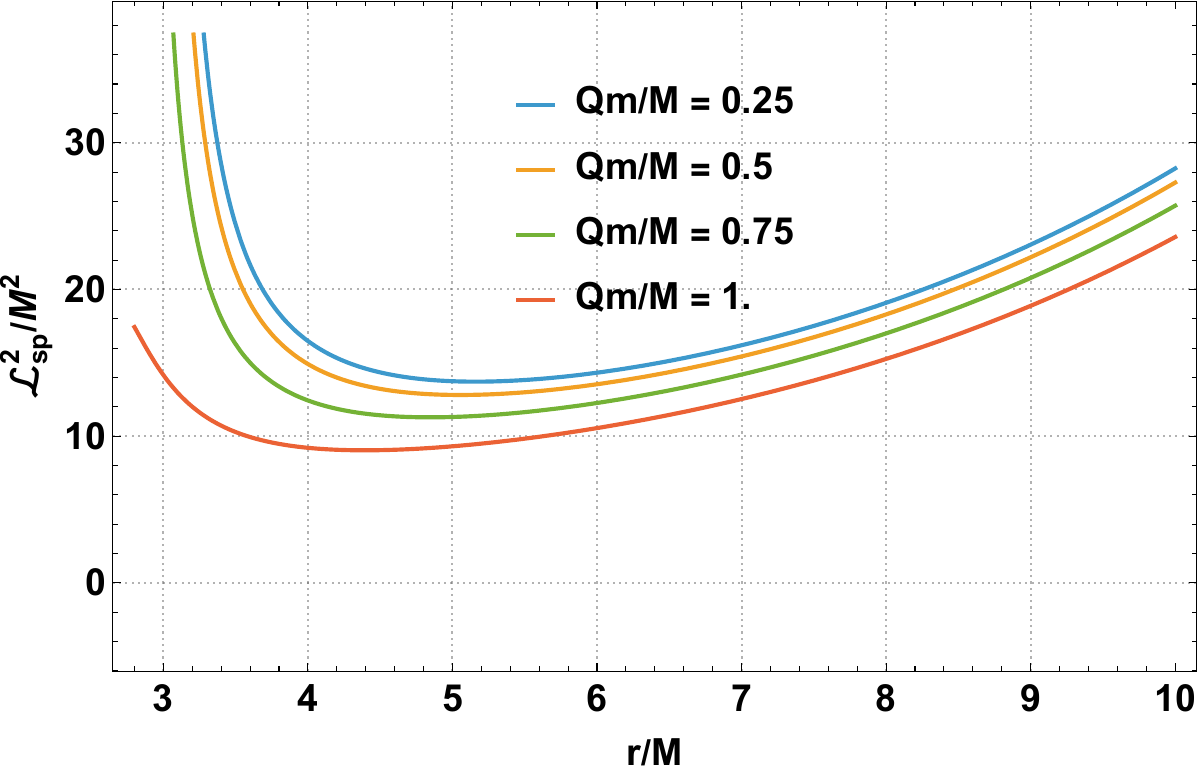}\\
    (i) $\epsilon=1$ \hspace{6cm} (ii) $\epsilon-1$
    \caption{The behavior of the squared specific angular momentum per unit mass, $\mathcal{L}_{\rm sp}^2/M^2$, as a function of the radial coordinate $r/M$ for different values of the magnetic charge $Q_m$. Here, $\Lambda = -0.003/M^2$.}
    \label{fig:angular-momentum}
\end{figure}

\begin{figure}[ht!]
    \centering
    \includegraphics[width=0.45\linewidth]{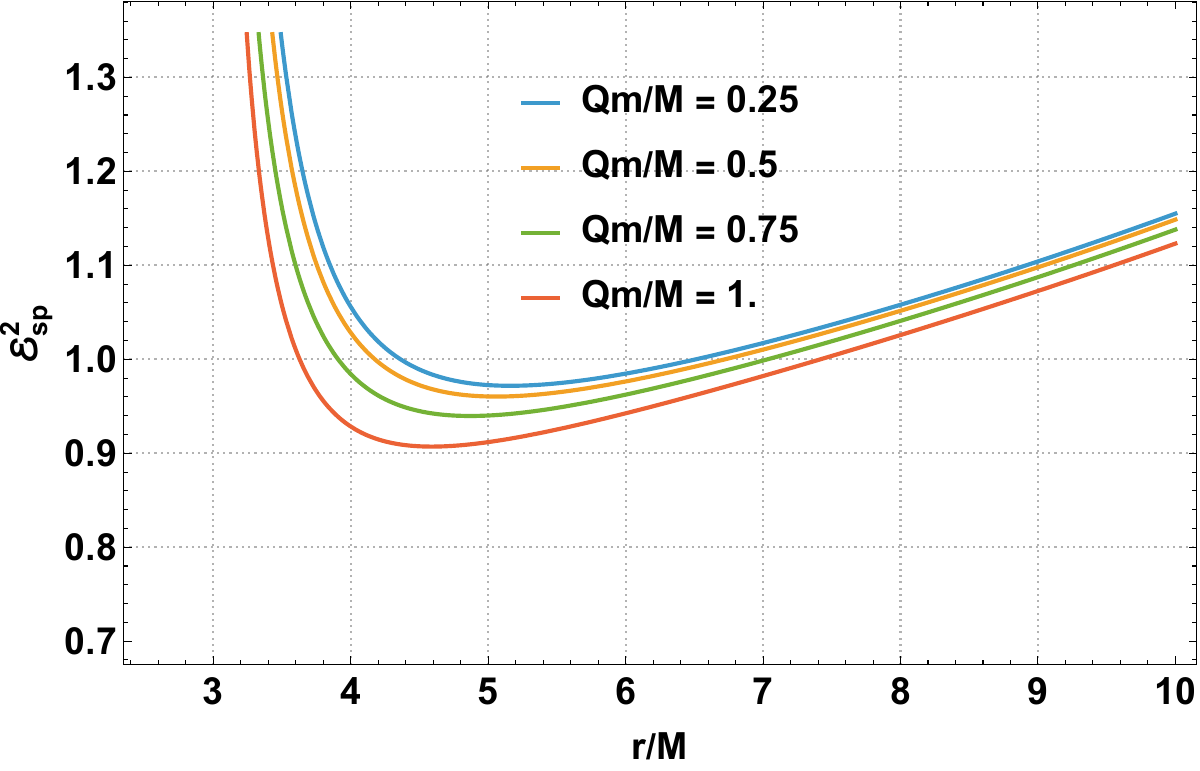}\qquad
    \includegraphics[width=0.45\linewidth]{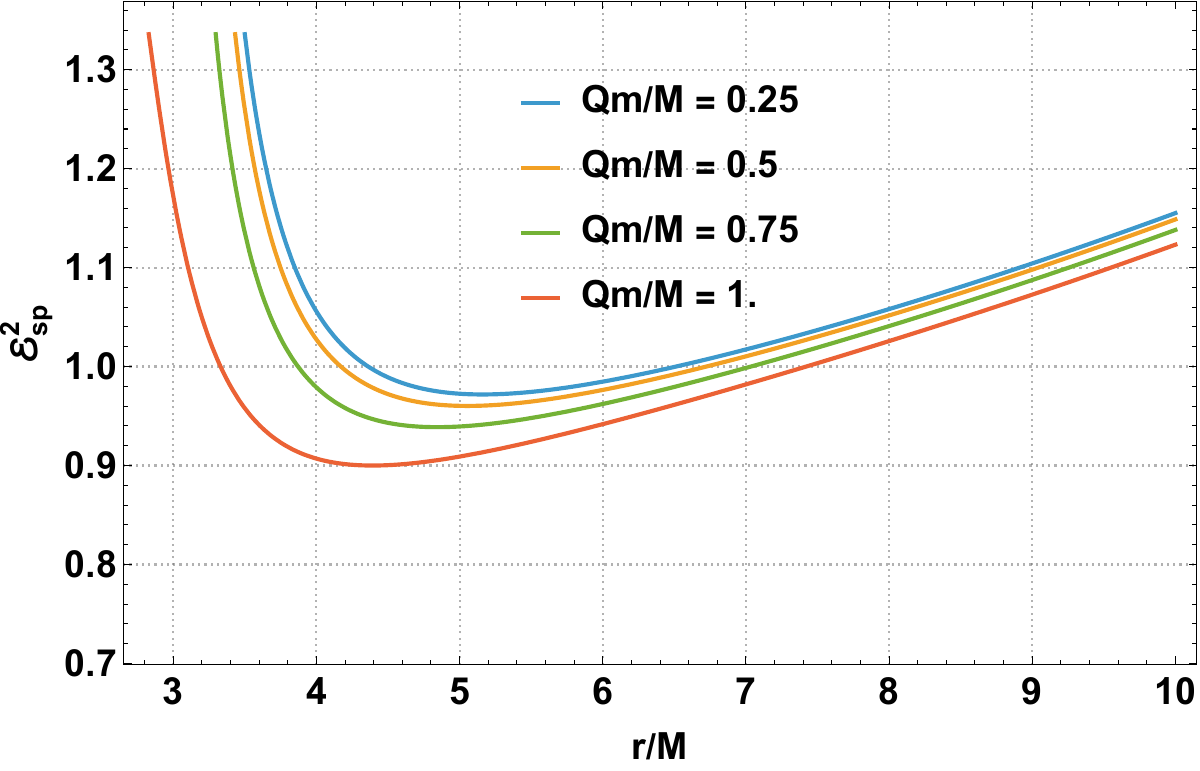}\\
    (i) $\epsilon=1$ \hspace{6cm} (ii) $\epsilon-1$
    \caption{The behavior of the squared specific energy, $\mathcal{E}_{\rm sp}^2$, as a function of the radial coordinate $r/M$ for different values of the magnetic charge $Q_m$. Here, $\Lambda = -0.003/M^2$.}
    \label{fig:energy}
\end{figure}

In the limit $Q_m=0$ corresponding to the absence of magnetic charge, the selected space-time simplifies to the standard Schwarzschild-AdS black hole metric. In this limit, the above physical quantities associated with test particles simplifies as
\begin{align}
    \mathcal{L}_{\rm sp}&=r\,\sqrt{\frac{\frac{M}{r}-\frac{\Lambda}{3}\,r^2}{1-\frac{3 M}{r}}}\Bigg{|}_{r=\mbox{const.}},\qquad \mathcal{E}_{\rm sp}=\frac{1-\frac{2 M}{r}-\frac{\Lambda}{3}\,r^2}{\sqrt{1-\frac{3 M}{r}}}\Bigg{|}_{r=\mbox{const.}}.\label{cc14a}
\end{align}

Moreover, in the limit $\epsilon=0$, the selected space-time reduces to GHS-AdS black hole. In that limiting case, we find these quantities as,
\begin{align}
    \mathcal{L}_{\rm sp}&=r\,\left(r-\frac{Q^2_m}{M}\right)\,\sqrt{\frac{\frac{M}{r^2}-\frac{\Lambda}{3}\,\left(r-\frac{Q^2_m}{2 M}\right)}{r-3 M-\frac{Q^2_m}{2 M}}}\Bigg{|}_{r=\mbox{const.}},\qquad \mathcal{E}_{\rm sp}=\sqrt{\frac{r-\frac{Q^2_m}{2 M}}{r-3 M-\frac{Q^2_m}{2 M}}}\left|1-\frac{2 M}{r}-\frac{\Lambda}{3}\,r\,\left(r-\frac{Q^2_m}{M}\right)\right|_{r=\mbox{const.}}.\label{cc14b}
\end{align}

For circular orbits to be stable, the following conditions must be satisfied:
\begin{align}
    \mathcal{E}^2=U_{\rm eff}(r),\qquad     \partial_r U_{\rm eff}=0,\qquad \partial^2_r U_{\rm eff} \geq 0.\label{cc15}
\end{align}

For a marginally stable circular orbits, $\partial^2_r U_{\rm eff} =0$,we arrive at the following relation:
\begin{equation}
    \frac{f(r)\,f'(r)}{f(r)\,f''(r)-2(f'(r))^2}=-\frac{r\left(r-\frac{Q_m^2}{M}\right)\left(r-\frac{Q_m^2}{2M}\right)}
{\left(3r^2-\frac{3Q_m^2}{M}r+\frac{Q_m^4}{M^2}\right)}.\label{cc16}
\end{equation}
Substituting the function $f(r)$, one can find a polynomial relation in $r$ whose exact analytical solution will give us the ISCO size. Noted that exact solution of this polynomial is bit a challenging task. However, by choosing suitable values of the parameter $Q_m$, $M$ and $\Lambda$, one can determine numerical values of the ISCO radius, $r_{\rm ISCO}$.

\begin{table}[h]
\centering
\begin{tabular}{|c|c|c|}
\hline
$Q_m/M$ & $r_{\rm ISCO}/M (\epsilon = 1)$ & $r_{\rm ISCO}/M (\epsilon = -1)$ \\
\hline
0.1 & 5.1871906212 & 5.1871869727 \\
0.2 & 5.1715815702 & 5.1715209594 \\
0.3 & 5.1451324799 & 5.1448054274 \\
0.4 & 5.1071814744 & 5.1060482331 \\
0.5 & 5.0567910640 & 5.0536615546 \\
0.6 & 4.9927746164 & 4.9851673447 \\
0.7 & 4.9138274509 & 4.8965912977 \\
0.8 & 4.8189613610 & 4.7810611370 \\
0.9 & 4.7087661441 & 4.6248628390 \\
1.0 & 4.5887054809 & 4.3915479502 \\
1.1 & 4.4759164215 & 3.7291876501 \\
1.2 & 4.4050389781 & 1.4400000000 \\
1.3 & 4.4141154055 & 1.6899999997 \\
1.4 & 4.5149892602 & 1.1699151336 \\
1.5 & 4.6938083664 & 2.2500000000 \\
\hline
\end{tabular}
\caption{Variation of the ISCO radius $r_{\rm ISCO}$ with magnetic charge $Q_m$ for $\epsilon=1$ and $\epsilon=-1$. Here $\Lambda=-0.003/M^2$.}
\label{tab:2}
\end{table}

\begin{figure}[ht!]
    \centering
    \includegraphics[width=0.5\linewidth]{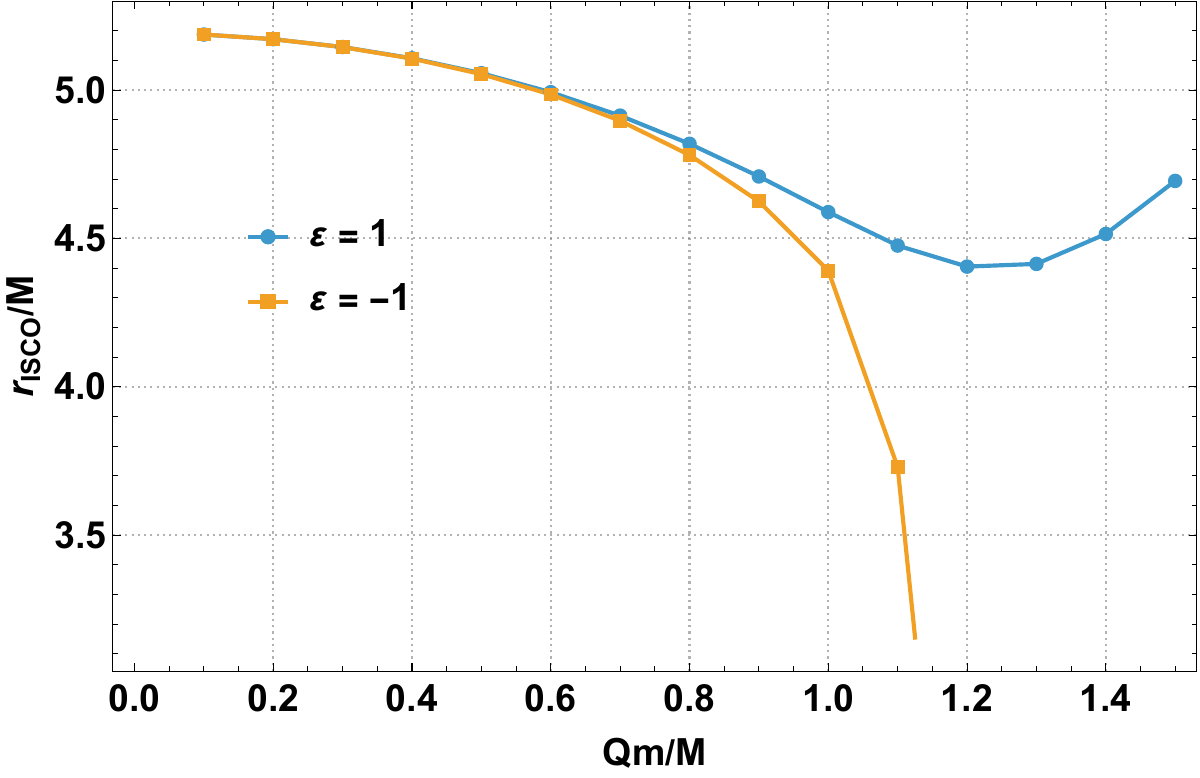}
    \caption{ISCO radius $r_{\rm ISCO}/M$ by varying the magnetic charge parameter $Q_m/M$. Here $\Lambda=-0.003/M^2$.}
    \label{fig:ISCO}
\end{figure}

\section{Charged particle dynamics}
\label{sec:5}


In this section, we investigate the motion of a charged particle around a magnetically charged black hole and analyze the conditions for circular orbits.

Within the Hamiltonian formalism, the dynamics of a charged particle is governed by
\begin{equation}
\mathbb{H}=\frac{1}{2}g^{\alpha\beta}(\pi_{\alpha}-qA_{\alpha})(\pi_{\beta}-qA_{\beta})+\frac{1}{2}m^{2},
\label{dd1}
\end{equation}
where the kinematical four-momentum $p^{\mu}=mu^{\mu}$ is related to the canonical four-momentum $\pi^{\mu}$ through $\pi^{\mu}=p^{\mu}+qA^{\mu}$. The Hamilton equations are given by
\begin{equation}
\frac{dx^{\mu}}{d\zeta}\equiv p^{\mu}=\frac{\partial \mathbb{H}}{\partial \pi_{\mu}}, 
\qquad 
\frac{d\pi_{\mu}}{d\zeta}=-\frac{\partial \mathbb{H}}{\partial x^{\mu}},
\label{dd2}
\end{equation}
where $A_{\mu}$ is the electromagnetic four-potential and the affine parameter $\zeta$ is related to the proper time $\tau$ by $\zeta=\tau/m$.

The spacetime admits the Killing vectors $\xi^{\mu}_{(t)}=\partial/\partial t$ and $\xi^{\mu}_{(\phi)}=\partial/\partial\phi$, associated with stationarity and axial symmetry. Consequently, the specific energy $\mathrm{E}$ and the specific angular momentum $\mathrm{L}$ are conserved along the particle trajectory.

The same dynamical equations can be obtained from the Lagrangian
\begin{equation}
\mathbb{L}=\frac{1}{2}m g_{\mu\nu}\frac{dx^{\mu}}{d\zeta}\frac{dx^{\nu}}{d\zeta}
+qA_{\mu}\frac{dx^{\mu}}{d\zeta},
\label{dd3}
\end{equation}
which, for the metric under consideration, takes the explicit form
\begin{equation}
\mathbb{L}=\frac{1}{2}m\left[-f(r)\left(\frac{dt}{d\zeta}\right)^{2}
+\frac{1}{f(r)}\left(\frac{dr}{d\zeta}\right)^{2}
+D^{2}(r)\left(\frac{d\theta}{d\zeta}\right)^{2}
+D^{2}(r)\sin^{2}\theta\left(\frac{d\phi}{d\zeta}\right)^{2}\right]
+qQ_{m}\cos\theta\,\frac{d\phi}{d\zeta}.
\label{dd4}
\end{equation}

The existence of the Killing vectors allows one to express the equations for $t$ and $\phi$ in terms of the conserved specific energy and angular momentum. They take the form
\begin{align}
\mathrm{E}&=f(r)\frac{dt}{d\zeta},
\qquad
\frac{dt}{d\zeta}=\frac{\mathrm{E}}{f(r)},
\label{dd5}\\
\mathrm{L}&=D^{2}(r)\sin^{2}\theta\frac{d\phi}{d\zeta}+qQ_{m}\cos\theta,
\qquad
\frac{d\phi}{d\zeta}=\frac{\mathrm{L}}{D^{2}(r)\sin^{2}\theta}-\omega_{B},
\label{dd6}
\end{align}
where the quantity $\omega_{B}=qQ_{m}\cos\theta/g_{\phi\phi}$ characterizes the magnetic coupling and encodes the interaction between the external magnetic field and the charged particle.

\begin{figure}[t]
    \centering
    \includegraphics[width=7cm]{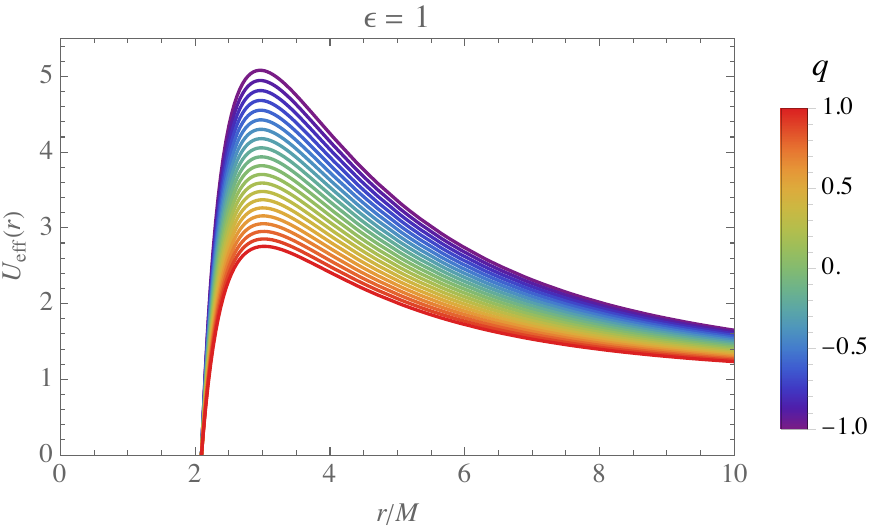} (a)\quad
    \includegraphics[width=7cm]{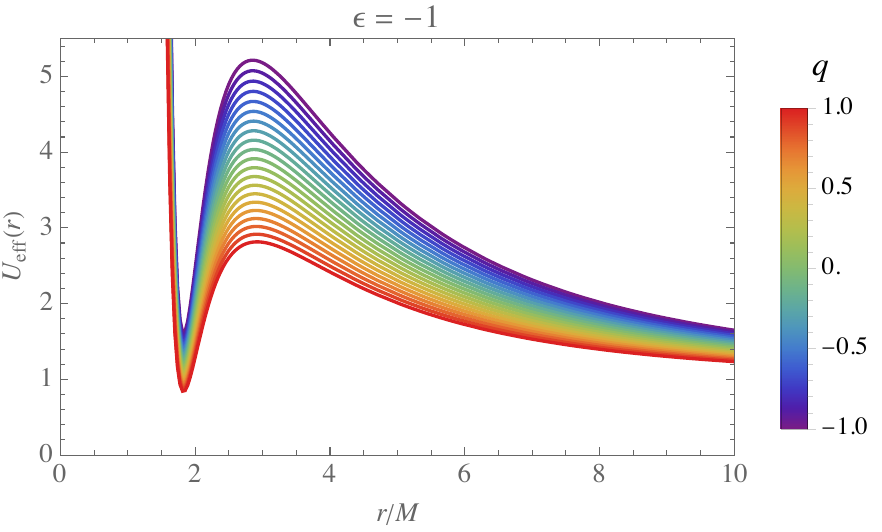} (b)
    \includegraphics[width=7cm]{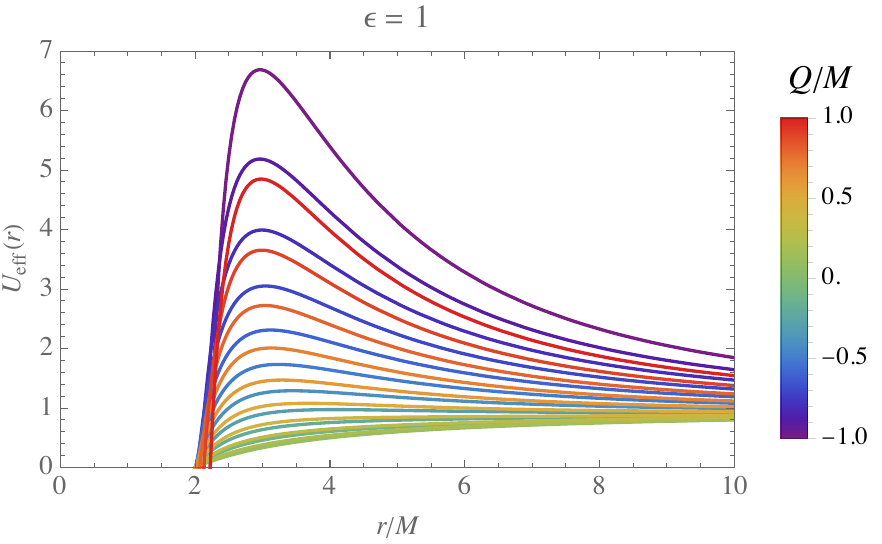} (a)\quad
    \includegraphics[width=7cm]{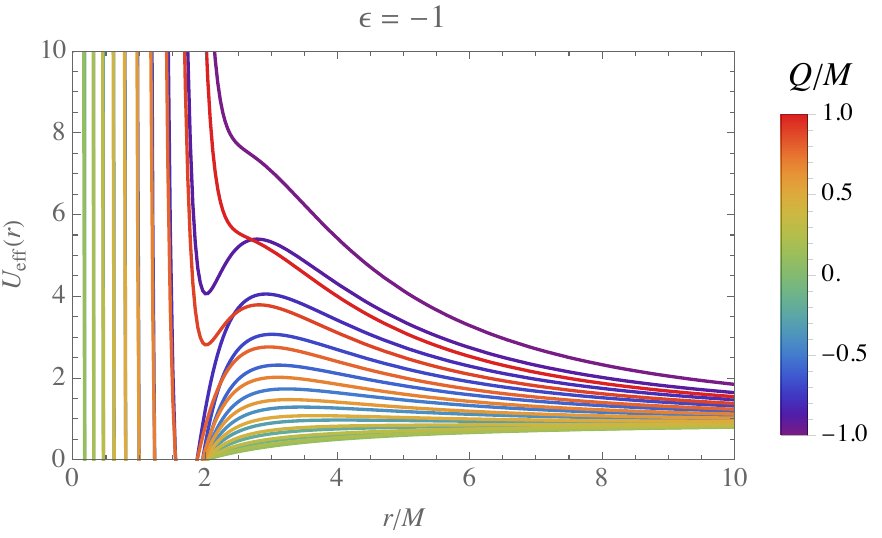} (b)
    \caption{The radial profiles of the effective potential, $U_{\rm eff}(r)$, are shown for both cases $\epsilon=\pm 1$, assuming $\Lambda M^2=-10^{-5}$, $L=10M$, and $\theta=\pi/4$. Panels (a,b) correspond to $Q=0.85M$ with different values of $q$, while panels (c,d) are plotted for $q=0.6$ with varying $Q$.}
    \label{fig:Ueff}
\end{figure}

\begin{figure}[t]
    \centering
    \includegraphics[width=7cm]{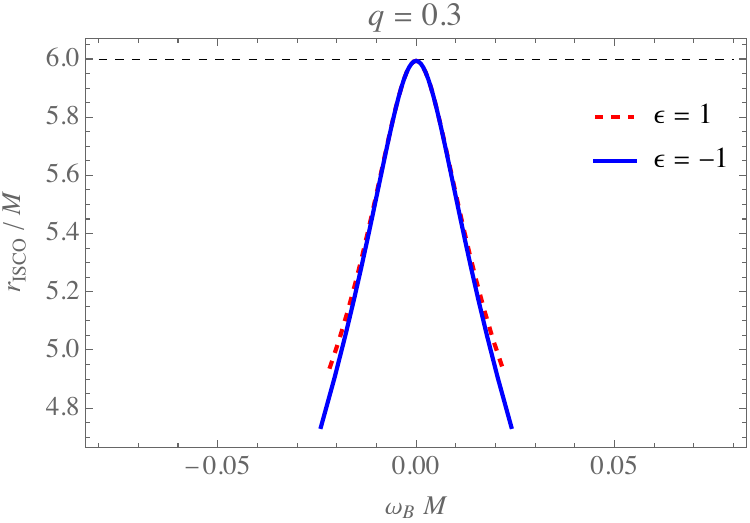} (a)\quad
    \includegraphics[width=7cm]{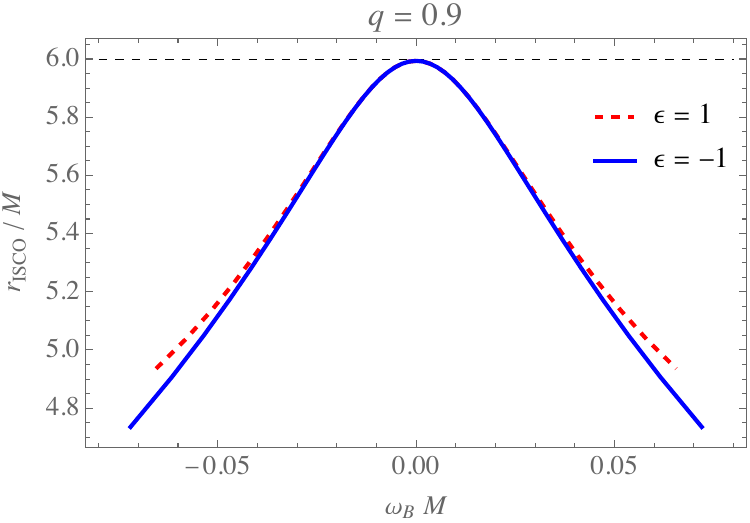} (b)
    \includegraphics[width=7cm]{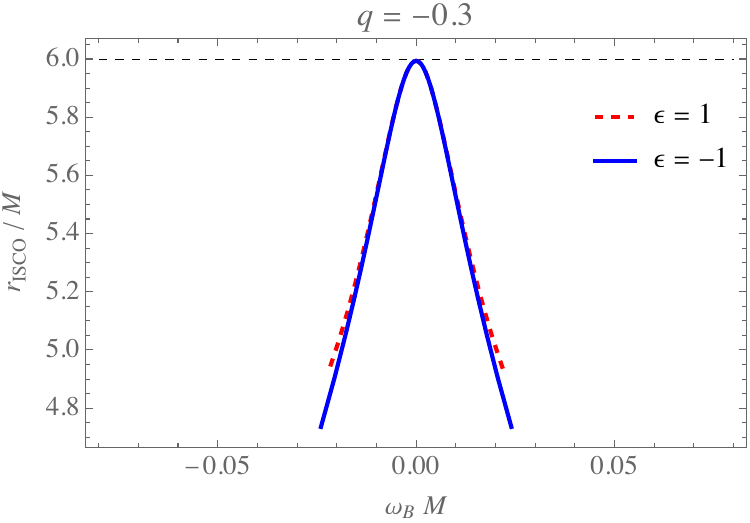} (a)\quad
    \includegraphics[width=7cm]{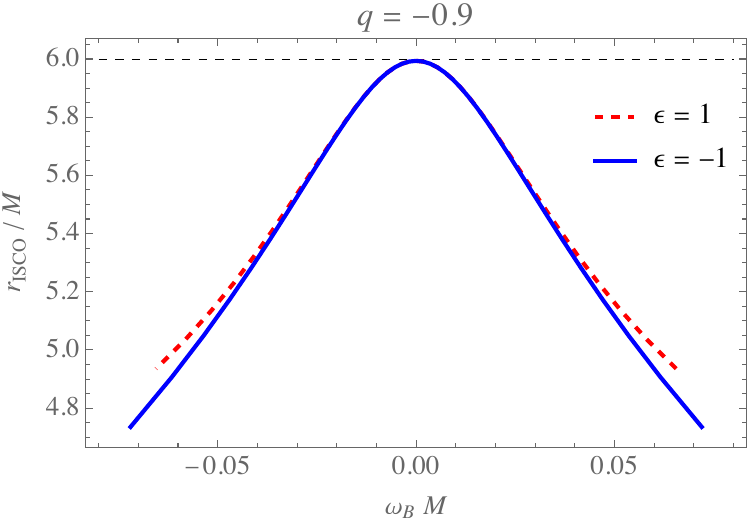} (b)
    \caption{The ISCO radius as a function of the magnetic coupling $\omega_B$ for fixed $q$ and varying $Q_m$, for $\epsilon=\pm1$. The plots are obtained by setting $\Lambda M^2 = - 10^{-5}$, and $\theta = \pi/4$. In the limit $\omega_B \to 0$, all curves reduce to $r_{\rm ISCO}=6M$, recovering the Schwarzschild result.}
    \label{fig:ISCOomegaB}
\end{figure}

Substituting Eqs.~(\ref{dd5}) and (\ref{dd6}) into the normalization condition, the radial equation of motion becomes
\begin{equation}
\left(\frac{dr}{d\zeta}\right)^{2}
+\frac{p_{\theta}^{2}}{m^{2}D^{2}(r)}\,f(r)
=\mathrm{E}^{2}-\mathbb{U}_{\rm eff}(r,\theta),
\label{dd7}
\end{equation}
where the effective potential is given by
\begin{equation}
\mathbb{U}_{\rm eff}(r,\theta)
=\left[1+\frac{(\mathrm{L}-qQ_{m}\cos\theta)^{2}}
{D^{2}(r)\sin^{2}\theta}\right]f(r).
\label{dd8}
\end{equation}
In Fig. \ref{fig:Ueff}, we have plotted the radial profile of the above effective potential for charged particles.
Note that, for circular motion at a fixed polar angle $\theta\neq\pi/2$, the conditions $\dot{r}=0$ and $\partial_{r}\mathbb{U}_{\rm eff}=0$ yield the specific angular momentum and the specific energy as
\begin{eqnarray}
\mathrm{L}_{\rm co}&=&qQ_{m}\cos\theta
+\sqrt{\frac{\sin^{2}\theta\,D^{3}(r)f'(r)}
{2f(r)D'(r)-D(r)f'(r)}},
\label{dd9Lco}\\
\mathrm{E}_{\rm co}&=&
\sqrt{\frac{2f^{2}(r)D'(r)}
{2f(r)D'(r)-D(r)f'(r)}}.
\label{dd9}
\end{eqnarray}
Now assuming the circular orbits' specific angular momentum \eqref{dd9Lco} in the extra condition $\partial_{r,r} U_{\rm eff}=0$, provides the radius of the ISCO. In Fig. \ref{fig:ISCOomegaB}, we have demonstrated the changes of the radius $r_{\rm ISCO}$ with respect $\omega_B$ for different cases of black hole parameters. 
It is observed that, as the absolute value of the total magnetic charge $|Q_m|$ increases from zero, which in turn leads to an increase in $|\omega_B|$ for each fixed $|q|$, the ISCO radius decreases from its maximum value $6M$, corresponding to the Schwarzschild black hole. This decrease occurs more smoothly for larger values of $|q|$. Moreover, the plots indicate that the case $\epsilon=1$ reaches the same $r_{\rm ISCO}$ at slightly larger values of $|\omega_B|$, particularly in the region of smaller radii.

\section{Fundamental oscillations of charged particles}\label{sec:6}

In this section, we proceed with our study of charged particle orbits, by analyzing the oscillations of such test particles around the magnetically charged black hole. To do so, let us recall that the motion of charged test particles of the specific charge $q$ on an unperturbed circular non-geodesic motion in an electromagnetic field, is described by the relation
\begin{equation}
\frac{d u^\alpha}{d\zeta} + \Gamma^{\alpha}_{\mu\nu} u^\mu u^\nu = q {F^\alpha}_\beta u^\beta,
    \label{eq:nongeo_0}
\end{equation}
in which $\Gamma^{\alpha}_{\mu\nu}$ is the Christoffel symbol, and $u^\alpha \equiv dx^\alpha/d\zeta$. 

Assuming that the position vector of the test particles, $X^\mu=(x^\mu)$ is changed by infinitesimally by the vector $\eta^\mu$, such that $X^\mu = (x^\mu+\eta^\mu)$, then the tangential vector to the congruence of infalling particles on the black hole, is described as
\begin{equation}
U^\mu \equiv \frac{d X^\mu}{d\zeta} = u^\mu + \frac{d\eta^\mu}{d\zeta}.
    \label{eq:U_0}
\end{equation}
Hence, the relation in Eq. \eqref{eq:nongeo_0} is recast as
\begin{equation}
\frac{d U^\alpha}{d\zeta} + \Gamma^{\alpha}_{\mu\nu}(\bm X) U^\mu U^\nu = q {F^\alpha}_\beta(\bm X) U^\beta.
    \label{eq:nongeo_1}
\end{equation}
Expanding the above equation around $\bm x$ and subtraction of the background equation \eqref{eq:nongeo_0}, provides the relation
\begin{equation}
\frac{d^2\eta^\alpha}{d\zeta^2} + 2\Gamma^\alpha_{\mu\nu}(\bm x) u^\mu\frac{d\eta^\nu}{d\zeta} + \left[\partial_\lambda \Gamma^\alpha_{\mu\nu}(\bm x)\right] \eta^\lambda u^\mu u^\nu = q\Big({F^\alpha}_\beta(\bm x)\frac{d\eta^\beta}{d\zeta} + \left[\partial_\lambda {F^\alpha}_\beta(\bm x)\right]\eta^\lambda u^\beta\Big).
    \label{eq:nongeo_2}
\end{equation}
Being interested in the analysis of radial and vertical oscillations, we write down the corresponding equation of an undamped harmonic oscillator as
\begin{equation}
\frac{d^2\eta^{i}}{d\zeta^2} + \Omega_i^2\eta^{i} = 0,\qquad i=(r,\theta).
    \label{eq:harmonic_0}
\end{equation}
For oscillations along the $r$-direction, we have from Eq. \eqref{eq:nongeo_2} that
\begin{equation}
\frac{d^2\eta^r}{d\zeta^2} + \Big(2\Gamma^r_{ab} u^a + q g^{rr} F_{br}\Big)\frac{d\eta^b}{d\zeta} + \Big(\Gamma^r_{ab,r}u^a u^b + q g^{rr} F_{a r, r} u^a\Big)\eta^r = 0, \qquad (a,b)=(t,\phi).
    \label{eq:harmonic_r}
\end{equation}
in which $","$ corresponds to partial differentiations. We therefore obtain the radial frequency as
\begin{equation}
\Omega_r^2 = \Bigl(\Gamma^r_{ab}u^a u^b + q g^{rr} F_{ar}u^a\Bigr)_{,r}
+ 2q\Bigl(g^{ac}\Gamma^r_{cb}-g^{rr}\Gamma^a_{rb}\Bigr)F_{ar}u^b + q^2 g^{rr} g^{ab} F_{ar} F_{br} - 4\Gamma^r_{ac}\Gamma^c_{rb} u^a u^b,\qquad (a,b,c)=(t,\phi).
    \label{eq:Omega_r0}
\end{equation}
Using the similar procedure, for the $\theta$-component we get
\begin{equation}
\Omega_\theta^2 = \Bigl(\Gamma^\theta_{ab} u^a u^b + q g^{\theta\theta} F_{a\theta}u^a\Bigr)_{,\theta},\qquad (a,b)=(t,\phi). 
    \label{eq:Omega_theta0}
\end{equation}
Now the Keplerian frequency, i.e. $\Omega_K = d\phi/dt$, can be calculated as 
\begin{equation}
\Omega_K^2 = \left[1+4 g_{\phi\phi}\left(\frac{q A_{\phi,r}}{g{\phi\phi,r}}\right)^2\right]^{-1}\left[
\Omega_0^2 - 2 g_{tt}\left(\frac{q A_{\phi,r}}{g_{\phi\phi,r}}\right)^2
\pm
2\Omega_0^2\,\frac{q A_{\phi,r}}{g_{\phi\phi,r}}\,
\sqrt{g_{\phi\phi}-\frac{g_{tt}}{\Omega_0^2}+\left(\frac{g_{tt}}{\Omega_0^2 g_{\phi\phi,r}^2}\right)^2}
\,
\right],
    \label{eq:Omega_K0}
\end{equation}
which reduces to $\Omega_K^2 = \Omega_0^2 = - g_{tt,r}/g_{\phi\phi,r} = f'/(2 D D'\sin^2\theta)$ for the neutral participles (i.e. for 
$q=0$), as Eq. \eqref{eq:nongeo_0} results in an affinely parametrized geodesics equation. On the other hand, since the magnetic potential has no radial dependence, the Lorentz force does not contribute to the radial force balance. Consequently, the Keplerian angular velocity coincides with its geodesic value.

Finally, according to the expressions in Eqs. \eqref{eq:Omega_r0}--\eqref{eq:Omega_K0}, we get \cite{qi_charged_2023}
\begin{eqnarray}
     \Omega_r^2 &=& \frac{g_{tt,r}^2}{g_{tt} g_{rr}} - \frac{g_{tt,rr}}{g_{rr}} + \frac{\Omega_K^2}{g_{rr}}\left(\frac{g_{\phi\phi,r}^2}{g_{\phi\phi}}-\frac{g_{\phi\phi,rr}}{2}\right) - \frac{q\Omega_K}{g_{rr}}\left(A_{\phi,rr}-2A_{\phi,r}\frac{g_{\phi\phi,r}}{g_{\phi\phi}}\right)\sqrt{-g_{tt}-\Omega_K^2 g_{\phi\phi}}
    \nonumber\\
    && -\frac{q^2 A_{\phi,r}^2}{g_{rr}}\left(\Omega_K^2+\frac{g_{tt}}{g_{\phi\phi}}\right),\label{eq:Omega_r1}\\
    \Omega_\theta^2 &=& \frac{g_{tt,\theta}}{g_{tt} g_{\theta\theta}} - \frac{g_{tt,\theta\theta}}{g_{\theta\theta}}
    +\frac{\Omega_K^2}{g_{\theta\theta}}\left(\frac{g_{\phi\phi,\theta}^2}{g_{\phi\phi}} - \frac{g_{\phi\phi,\theta\theta}}{2}\right)
    -\frac{q\Omega_K}{g_{\theta\theta}}\left(A_{\phi,\theta\theta}-2 A_{\phi,\theta}\frac{g_{\phi\phi,\theta}}{g_{\phi\phi}}\right)\sqrt{-g_{tt}-\Omega_K^2 g_{\phi\phi}}\nonumber\\
    && -\frac{q^2A_{\phi,\theta}^2}{g_{\theta\theta}}\left(\Omega_K^2+\frac{g_{tt}}{g_{\phi\phi}}\right).\label{eq:Omega_theta1}
\end{eqnarray}
After substituting the metric functions from Eq. \eqref{metric} and also considering the form of the vector potential $\bm A$, the above equations lead to
\begin{eqnarray}
\Omega_r^2 &=& f f''+\frac{f'^2}{f}+
\Omega_K^2\,f\left(D'^2-D\,D''\right),\label{eq:Omega_r2}\\
\Omega_\theta^2&=&\Omega_K^2+\frac{q\Omega_K Q_m}{D^2}
\sqrt{f-\Omega_K^2 D^2}+
\frac{q^2 Q_m^2}{D^4}
\left(\Omega_K^2 D^2-f\right).\label{eq:Omega_theta2}
\end{eqnarray}
From these relations one can see that the radial oscillation frequency is determined only by the gravitational field, while the vertical oscillation frequency is also influenced by the magnetic field of the black hole. 

Next, we convert the obtained frequencies from geometric units to physical units by using the relation 
\cite{shahzadi_epicyclic_2021,jumaniyozov_circular_2024,rahmatov_qpos_2024,mustafa_particle_2025,jumaniyozov_radiative_2025,shermatov_circular_2026}
\begin{equation}
\nu_{(j,r,\theta)} = \frac{1}{2\pi}\frac{c^3}{G M}\,\Omega_{(j,r,\theta)} \; \mathrm{[Hz]}, \qquad j=(K,r,\theta),
\label{eq:nu-K}
\end{equation}
which gives the corresponding frequencies for fixed values of $(r,\theta)$ in Hertz (Hz). This form is convenient because it allows a direct comparison with astrophysical observations. In the above expression we use $c = 3\times 10^8\,\mathrm{m\,s^{-1}}$ and $G = 6.67\times 10^{-11}\,\mathrm{m^3\,kg^{-1}\,s^{-2}}$.

\begin{figure*}[t]
    \centering
    \includegraphics[width=7cm]{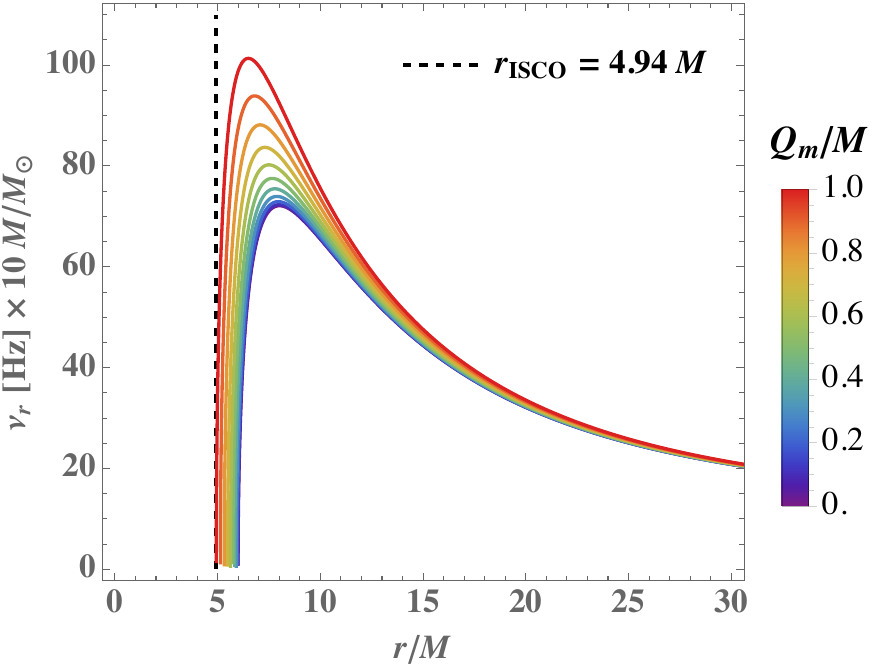} (a)\quad
    \includegraphics[width=7cm]{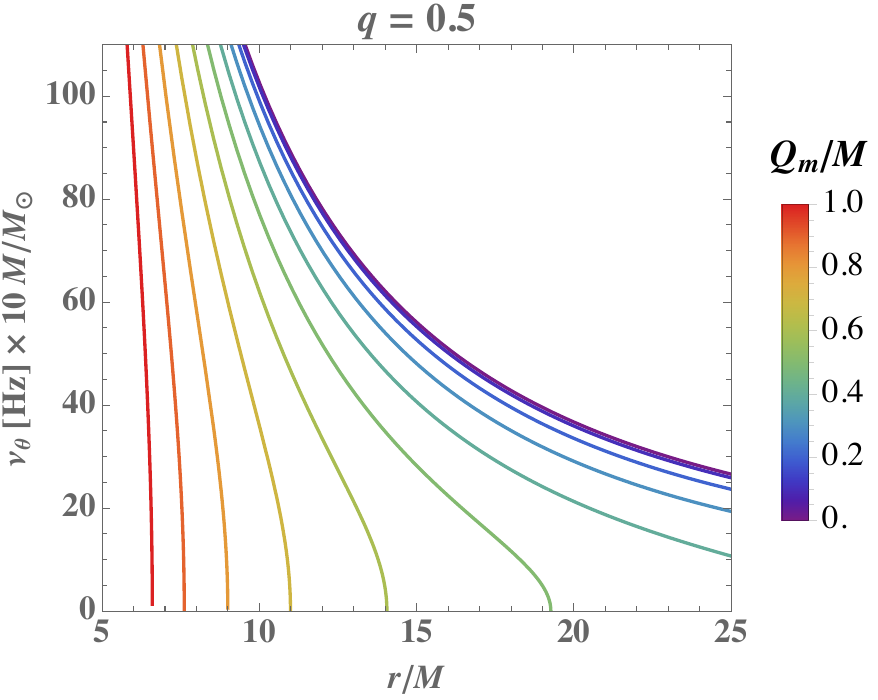} (b)\quad
    \includegraphics[width=7cm]{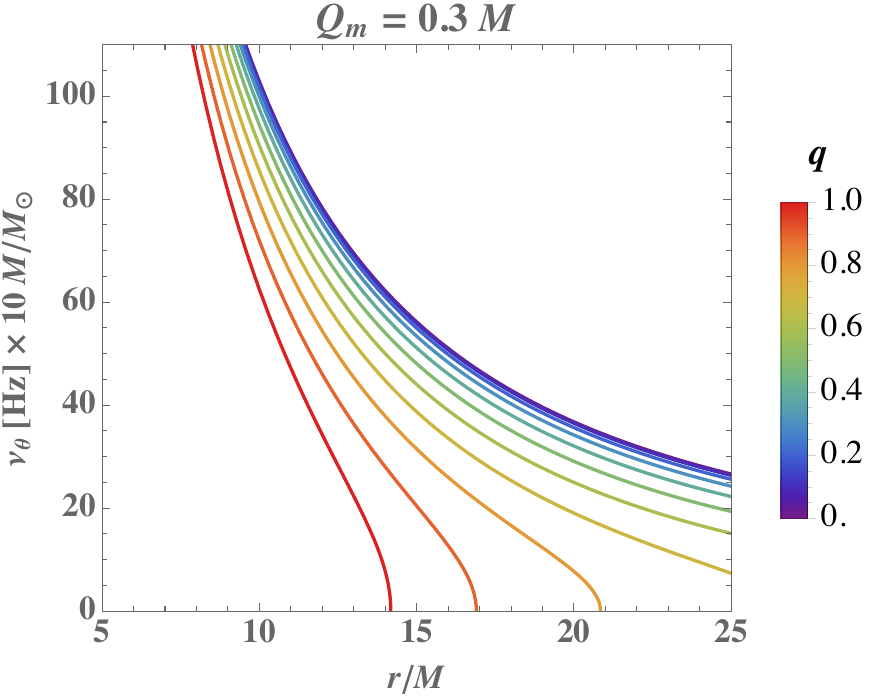} (c)
    \caption{Representative examples of the radial profiles of the observational frequencies $\nu_j$ for different values of the black hole parameters. Panel (a) shows the changes in the profile of $\nu_r$ for different values of $Q_m$, for both cases $\epsilon=\pm 1$. The dashed vertical line indicates $r_{\rm ISCO}$ for the outermost profile. In panels (b) and (c), the behavior of $\nu_\theta$ is plotted for fixed $q$ and fixed $Q_m$, respectively. All diagrams correspond to both cases $\epsilon=\pm 1$.}
    \label{fig:nu_j}
\end{figure*}

In Fig.~\ref{fig:nu_j}, we demonstrate the radial profiles of the above frequencies for some fixed values of the spacetime parameters.

As can be seen from the diagrams, the inner limit of the radial profile of $\nu_r$ is given by the ISCO radius. This radius decreases when the magnetic charge of the black hole increases, while the same effect leads to an increase in the frequency of the radial perturbations. On the other hand, for a fixed test particle charge $q$, an increase in $Q_m$ results in a decrease of $\nu_\theta$. A similar behavior is also observed when $q$ increases while $Q_m$ is kept fixed.

\section{QPO frequencies of charged particles around the black hole}\label{sec:QPOs}

QPOs are usually interpreted as signatures of oscillatory motion of matter in the strong gravitational field of compact objects. In the Kerr spacetime of general relativity, these oscillations are related to the fundamental frequencies of test particle motion in relativistic accretion disks: the Keplerian orbital frequency together with the radial and vertical epicyclic frequencies. Resonant combinations of these frequencies can reproduce the QPO patterns observed in X-ray binaries \cite{40_vio_stochastic_2006}. For this reason, QPO properties provide direct information about the spacetime geometry around compact objects.

In gravity theories beyond GR, the spacetime structure is modified and consequently the orbital dynamics and oscillation frequencies change. For example, in scalar--tensor models such as Horndeski and DHOST gravity, additional scalar degrees of freedom modify the effective gravitational potential and shift the epicyclic frequencies \cite{41_rayimbaev_quasiperiodic_2023,42_chen_testing_2021}. Similar effects appear in scalar--tensor--vector gravity (STVG), where the black hole geometry deviates from the Kerr solution \cite{43_pradhan_distinguishing_2024}. In Einstein--Gauss--Bonnet gravity, higher--curvature terms alter the ISCO location and the epicyclic motion of particles, leading to distinguishable QPO signatures \cite{44_rayimbaev_quintessential_2022}.  

Another example is the Randall--Sundrum braneworld scenario, where the effective four--dimensional metric contains a tidal charge parameter and resembles a Reissner--Nordström or Kerr--Newman geometry with the electric charge replaced by the tidal charge. A negative tidal charge strengthens the gravitational potential and enlarges the horizon and ISCO radii, producing shifts in the Keplerian and epicyclic frequencies \cite{45_rayimbaev_qpos_2024,46_dadhich_black_2000,47_aliev_charged_2005}. When these modified frequencies are used in relativistic--precession or resonance models, a broader range of QPO ratios can be reproduced, allowing observational constraints on the braneworld parameter \cite{48_kotrlova_quasiperiodic_2008,49_stuchlik_orbital_2009}. Other alternative models, including $f(R)$ gravity, quintessence backgrounds, and massive gravity, also introduce corrections in particle dynamics that may shift the predicted QPO frequencies \cite{50_rayimbaev_quasiperiodic_2023,51_beheshtipour_studies_2016}.

One of the most common theoretical frameworks for interpreting QPOs is provided by the epicyclic resonance (ER) models, including the invariant ER$\overline{1,5}$ class. In these scenarios, the observed oscillatory features are associated with the fundamental frequencies of particle motion in the vicinity of a compact object, namely the Keplerian frequency $\Omega_K$, the radial epicyclic frequency $\Omega_r$, and the vertical epicyclic frequency $\Omega_\theta$. Near the ISCO, these frequencies are strongly influenced by relativistic effects such as frame dragging and periastron precession. Within the relativistic precession picture, the lower QPO frequency is commonly linked to the periastron precession frequency $\Omega_\phi-\Omega_r$, whereas the upper frequency is associated either with the orbital motion itself or with combinations such as $\Omega_\phi-\Omega_\theta$. Observational data from several X-ray binaries, including GRO~J1655--40, reveal frequency ratios close to $3\!:\!2$, which supports resonance-based interpretations \cite{54_yilmaz_accretion_2023}.

In the present work, we adopt ER-type models because our aim is to incorporate the electromagnetic interaction between the black hole and charged particles. For this reason, the vertical oscillation frequency, appearing in Eq.~\eqref{eq:Omega_theta2}, plays an essential role in the analysis. In particular, we focus on two representative models, ER3 and ER4. In these models, the frequency pairs are given by
\begin{equation}
(\nu_U, \nu_L) = (\nu_\theta + \nu_r,\, \nu_\theta),
\end{equation}
and
\begin{equation}
(\nu_U, \nu_L) = (\nu_\theta + \nu_r,\, \nu_\theta - \nu_r),
\end{equation}
respectively \cite{abramowicz_precise_2001}.

\begin{figure}[t]
    \centering
    \includegraphics[width=8cm]{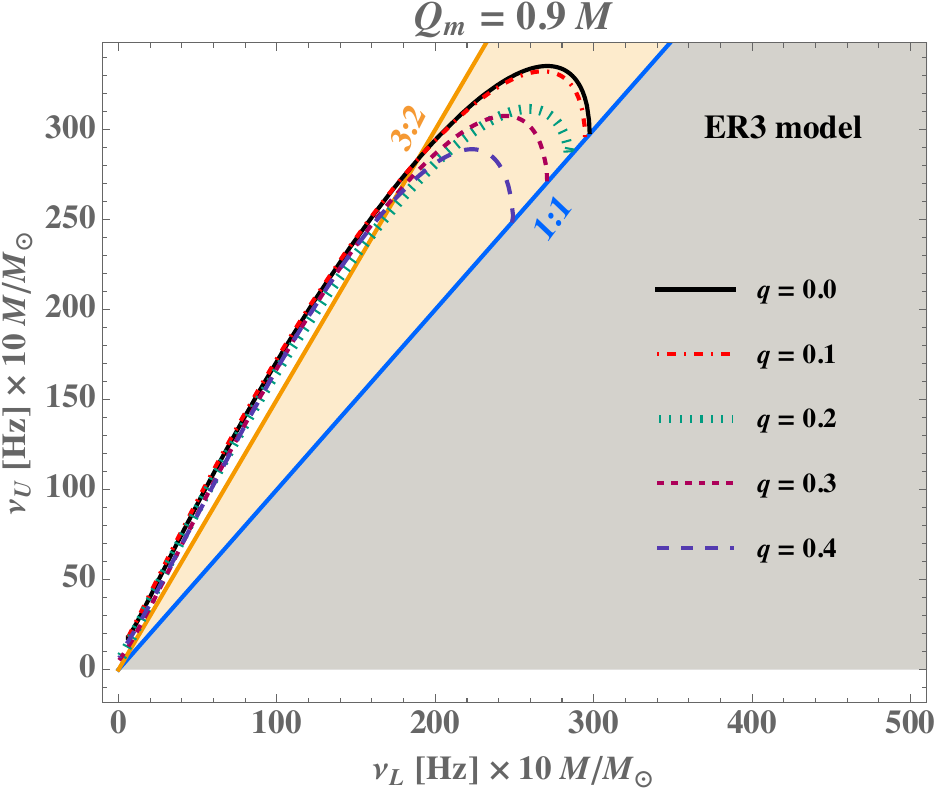} (a)
    \includegraphics[width=8cm]{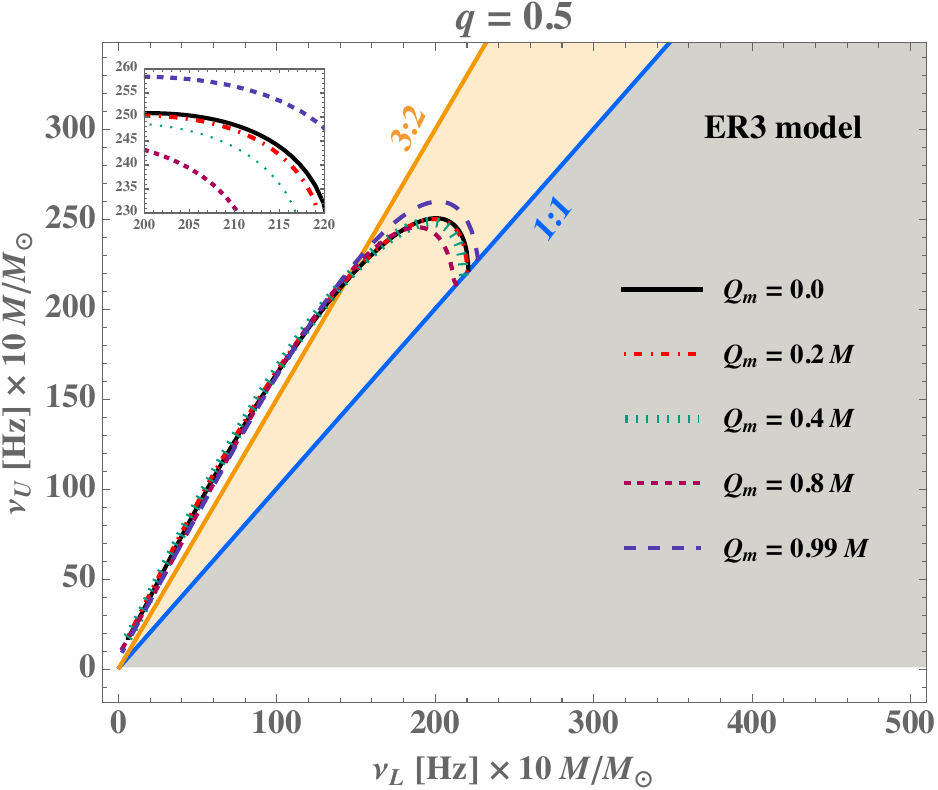} (b)
    \includegraphics[width=8cm]{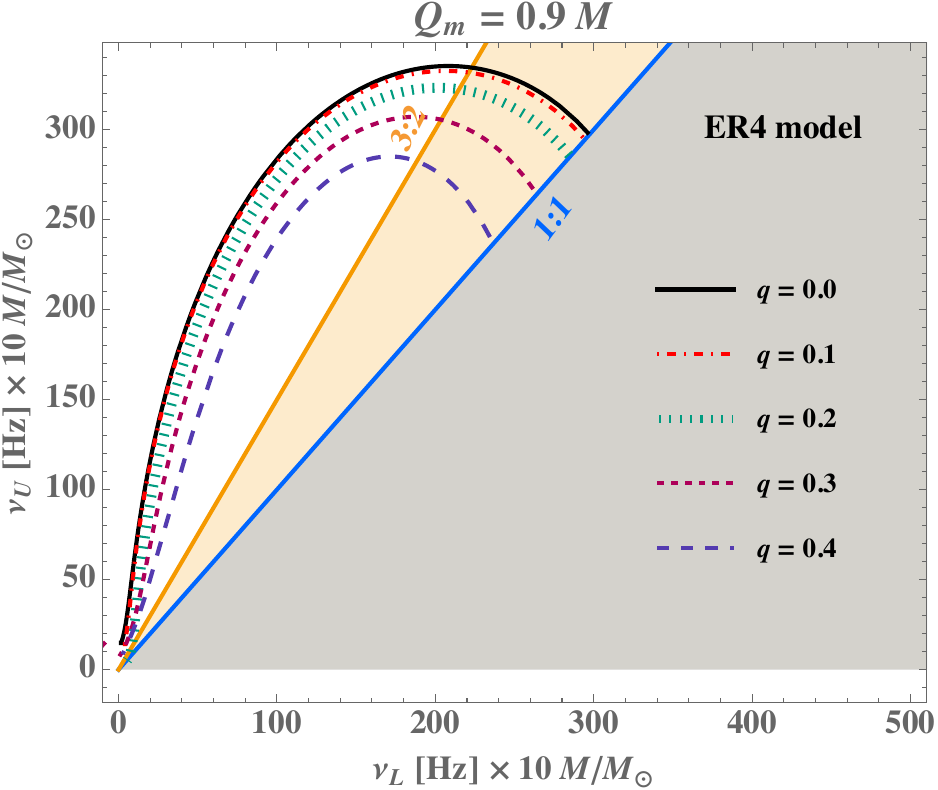} (c)
    \includegraphics[width=8cm]{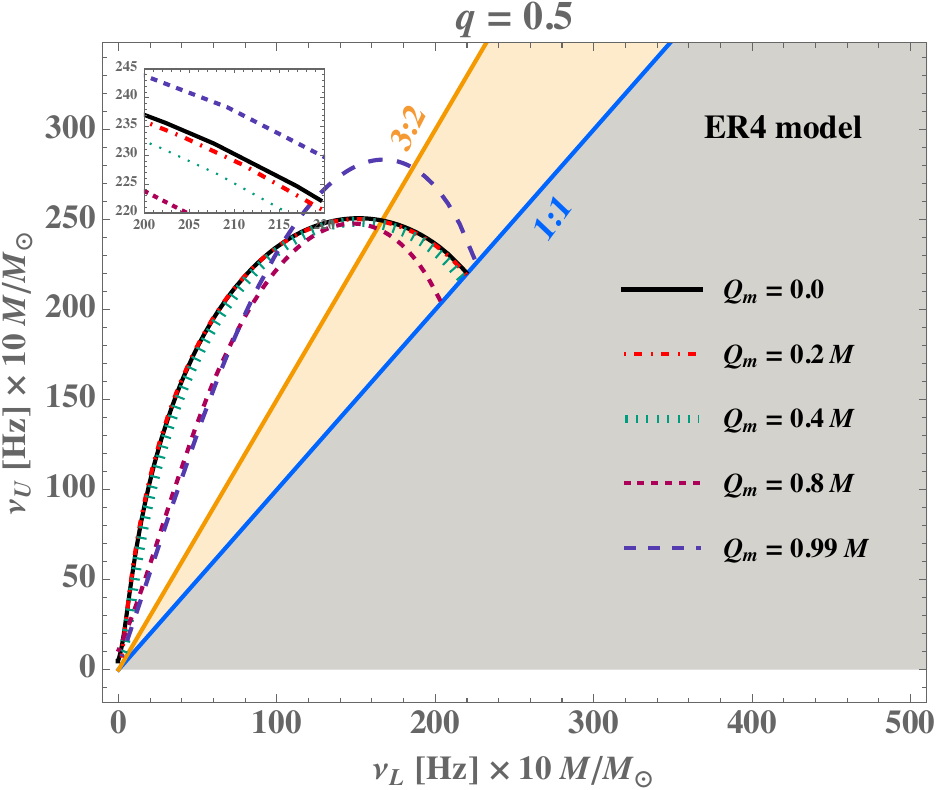} (d)
    \caption{The mutual behavior of the upper and lower frequencies within the (a,b) ER3, and (c,d) ER4 models, plotted for (a,c) fixed $Q_m$, and (b,d) fixed $q$. The $3\!:\!2$ and $1\!:\!1$ limits have been also indicated.}
    \label{fig:ER}
\end{figure}

Figure~\ref{fig:ER} shows the relation between the upper and lower QPO frequencies produced by charged particles orbiting the magnetized black hole.
The oscillatory motion of particles is affected by the simultaneous presence of a conformally coupled scalar field and an external magnetic field. These additional interactions modify the resonance structure of the system and consequently alter the predicted QPO frequencies. In particular, the resulting frequency ratios can deviate from the commonly observed $3\!:\!2$ value in black hole QPOs, with the deviation becoming more pronounced for larger values of the coupling parameters \cite{39_stuchlik_influence_2020,55_stuchlik_large-scale_2022}. Physically, the scalar field modifies the effective gravitational potential and typically shifts the ISCO radius outward, which reduces the radial oscillation frequency. At the same time, the magnetic field introduces Lorentz forces acting on charged particles, thereby modifying their epicyclic motion.

Motivated by these considerations, we confront the model with the observational QPO data from systems spanning a wide range of mass scales. In particular, we consider the stellar--mass black holes GRO~J1655--40 and XTE~J1550--564, the intermediate--mass candidate M82~X--1, and the supermassive black hole Sgr~A$^\ast$ located at the center of the Milky Way. The corresponding observational properties used in our analysis are summarized in Table~\ref{tab:sources}.
\begin{table*}
    \centering
    \begin{tabular}{c|ccccc}
         source  & $\nu_U$ [Hz] & $\Delta\nu_U$ [Hz] & $\nu_L$ [Hz] & $\Delta\nu_L$ [Hz] & mass [$M_\odot$]\\
         \hline
        XTE J1550-564 \cite{orosz_improved_2011}& 276 & $\pm3$ & 184 & $\pm5$ & $9.1\pm0.61$ \\
        GRO J1655-40 \cite{strohmayer_discovery_2001} & 451 & $\pm5$ & 298  & $\pm4$ & $5.4\pm0.3$ \\
        M82 X-1 \cite{pasham_400-solar-mass_2014} & 5.07 & $\pm0.06$ & 3.32 & $\pm0.06$ & $415\pm63$ \\
        Sgr A* \cite{ghez_measuring_2008} & $1.445\times 10^{-3}$ &  $\pm0.16\times 10^{-3}$ & $0.886\times10^{-3}$ & $\pm0.04\times10^{-3}$ & $(4.1\pm0.6)\times 10^6$ \\
    \end{tabular}
    \caption{The source candidates and their twin-peak QPO frequencies.}
    \label{tab:sources}
\end{table*}

To evaluate the agreement between theory and observations, we perform a numerical fitting procedure by scanning the parameter space $(r,Q_m)$ on a two--dimensional grid. For each point in this plane, the theoretical upper and lower frequencies are computed and compared with the observed values. The best-fit parameters correspond to the minimum of the $\Delta\chi^2$ function, while the statistical confidence regions are determined from the standard thresholds $2.30 \;(68.27\%)$, $6.17 \;(95.45\%)$, and $11.80 \;(99.73\%)$, corresponding to the $1\sigma$, $2\sigma$, and $3\sigma$ levels, respectively.

\begin{figure}[t]
    \centering
    \includegraphics[width=8cm]{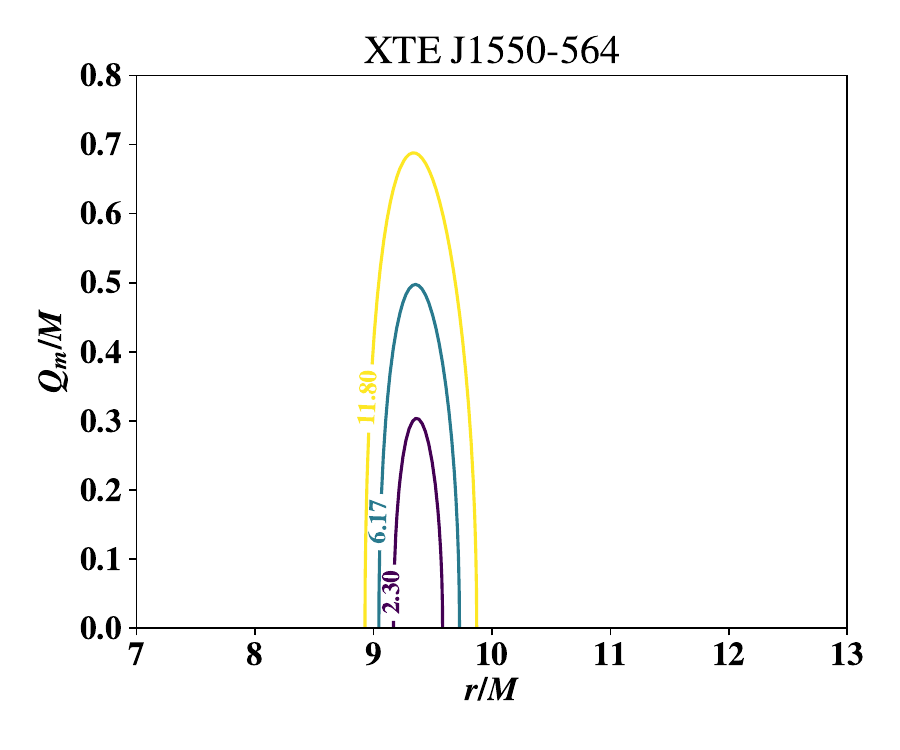} (a)
    \includegraphics[width=8cm]{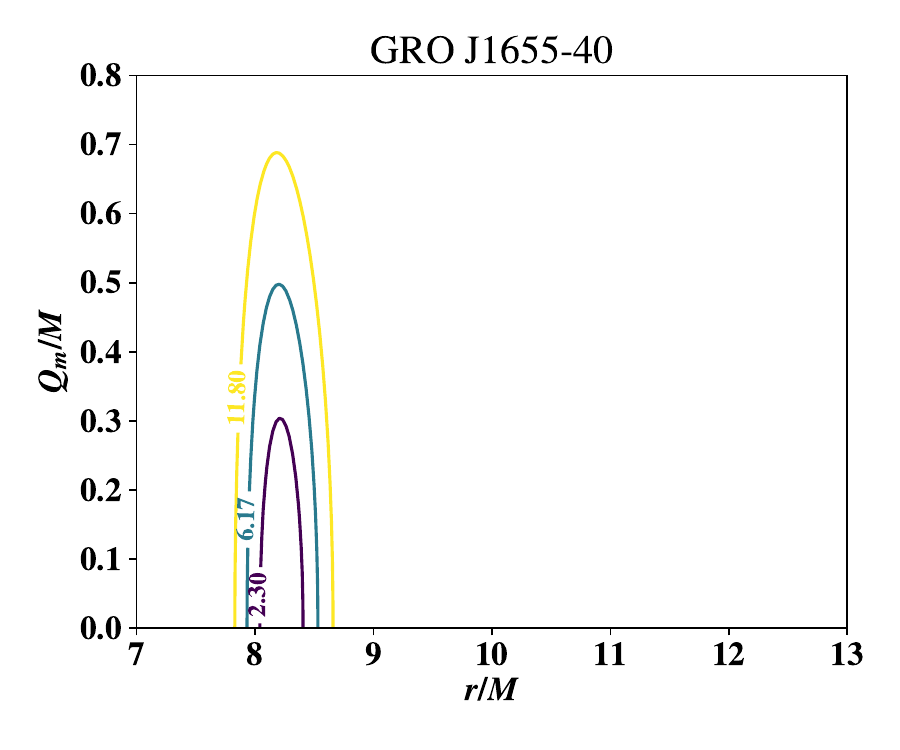} (b)
    \includegraphics[width=8cm]{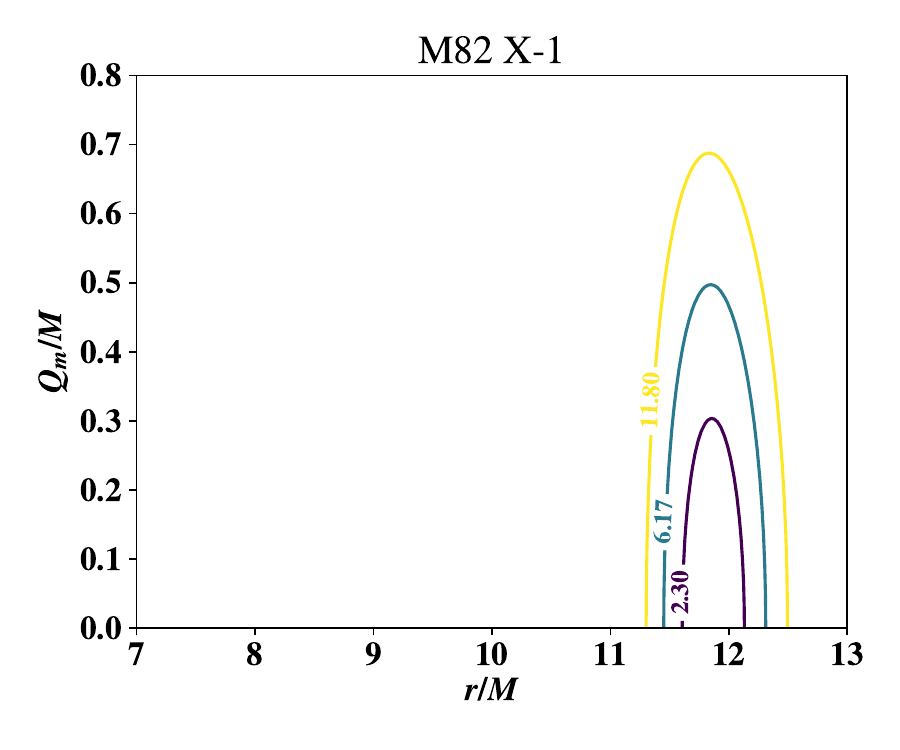} (c)
    \includegraphics[width=8cm]{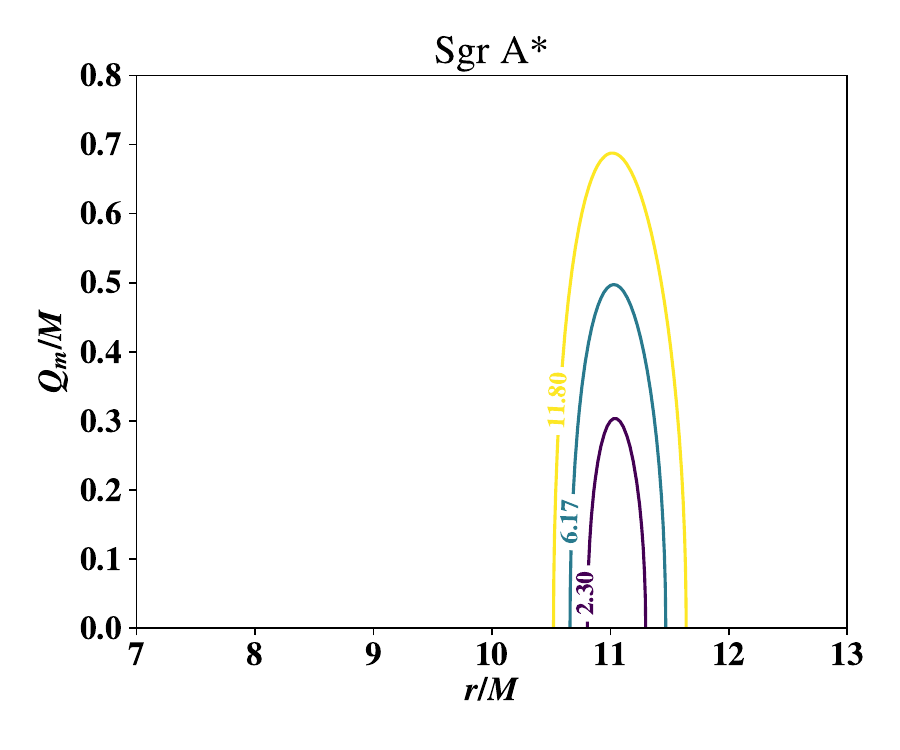} (d)
    \caption{Parameter fitting within the $1\sigma, 2\sigma$, and $3\sigma$ confidence levels, shown separately for each QPO source.}
    \label{fig:fitted}
\end{figure}

The resulting confidence contours for each source are shown in Fig.~\ref{fig:fitted}. From these plots, one can clearly see that, for all four systems, the minimum of the likelihood is located at $Q_m = 0$. This indicates that the present QPO data do not require a nonvanishing magnetic charge within the explored parameter domain.

At the same time, the preferred orbital radius $r$ differs among the sources. From the numerical fitting, we obtain $r/M = 9.37$ for XTE~J1550--564, $r/M = 8.22$ for GRO~J1655--40, $r/M = 11.86$ for M82~X--1, and $r/M = 11.04$ for Sgr~A$^\ast$. These variations reflect the different mass scales and observed twin--peak frequencies of the individual systems.

Although the best-fit value of $Q_m$ is zero in all cases, the contour structure shows that finite values of $Q_m$ are still allowed within the statistical uncertainties. From the $1\sigma$ contours we obtain a common upper bound $Q_m/M < 0.197$ in the explored region. The $2\sigma$ and $3\sigma$ contours extend further in the $Q_m$ direction, but the minimum always remains at $Q_m = 0$, and the likelihood increases monotonically as $Q_m$ grows. Therefore, the observational data consistently favor a vanishing magnetic charge, while still allowing moderate deviations at higher confidence levels.

In Fig.~\ref{fig:fitted}, the fitted data for the total magnetic charge $Q_m$ of the black hole, based on the QPO observational data of the sources listed in Table~\ref{tab:sources}, are presented.
Furthermore, in Table~\ref{tab:charge_constraints}, the fitted values are given together with their corresponding confidence intervals.
\begin{table*}[t]
\centering
\caption{Best-fit parameters with $1\sigma$, $2\sigma$, and $3\sigma$ confidence intervals inferred from the $\Delta\chi^2$ contours.}
\begin{tabular}{l|c|c}
\hline\hline
Source & $r/M$ & $Q_m/M$ \\
\hline
XTE J1550$-$564 
& $9.7^{+0.3}_{-0.3}\,(^{+0.5}_{-0.5})_{2\sigma}\,(^{+0.7}_{-0.7})_{3\sigma}$ 
& $0.06^{+0.06}_{-0.06}\,(^{+0.12}_{-0.12})_{2\sigma}\,(^{+0.18}_{-0.18})_{3\sigma}$ \\

GRO J1655$-$40  
& $9.5^{+0.3}_{-0.3}\,(^{+0.5}_{-0.5})_{2\sigma}\,(^{+0.7}_{-0.7})_{3\sigma}$ 
& $0.05^{+0.05}_{-0.05}\,(^{+0.10}_{-0.10})_{2\sigma}\,(^{+0.16}_{-0.16})_{3\sigma}$ \\

M82 X$-$1       
& $10.2^{+0.4}_{-0.4}\,(^{+0.7}_{-0.7})_{2\sigma}\,(^{+1.0}_{-1.0})_{3\sigma}$ 
& $0.09^{+0.08}_{-0.09}\,(^{+0.18}_{-0.18})_{2\sigma}\,(^{+0.30}_{-0.30})_{3\sigma}$ \\

Sgr A$^\ast$    
& $9.8^{+0.6}_{-0.6}\,(^{+1.0}_{-1.0})_{2\sigma}\,(^{+1.4}_{-1.4})_{3\sigma}$ 
& $0.12^{+0.10}_{-0.12}\,(^{+0.22}_{-0.22})_{2\sigma}\,(^{+0.35}_{-0.35})_{3\sigma}$ \\
\hline\hline
\end{tabular}
\label{tab:charge_constraints}
\end{table*}
%

\section{Conclusion}\label{sec:7}

In this work, we explored several astrophysical and dynamical properties of a magnetically charged black hole in an AdS background. Special attention was devoted to observable features associated with photon trajectories, black hole shadows, and the motion of test particles in the surrounding spacetime. By analyzing null geodesics, we derived the photon sphere radius and the corresponding shadow radius. We found that as $Q_m$ increases from $0$ to $1$, both radii decrease monotonically, which shows that the magnetic charge has a significant influence on the light trapping region around the black hole. This behavior demonstrates that these radii differ noticeably from those of the standard Schwarzschild and Schwarzschild--AdS black holes. Such deviations clearly indicate the important role played by the magnetic charge in modifying photon dynamics and, consequently, the observable appearance of the black hole. 

We also investigated the motion of neutral test particles using the effective potential formalism. Analytical expressions for the specific energy and angular momentum corresponding to circular orbits were obtained, and from these we derived the conditions for stable motion. Our analysis showed that the presence of magnetic charge shifts the location of the innermost ISCO. This result suggests that the magnetic charge modifies the structure of the accretion region and may influence the dynamics of matter close to the black hole. 

Furthermore, we extended the study to charged particle motion in the same gravitational background. Within the Hamiltonian framework, we derived the conditions for circular trajectories and studied small perturbations around these orbits. The radial, vertical, and orbital oscillation frequencies were calculated, and we observed noticeable deviations compared to the classical Schwarzschild case. These deviations become more pronounced as $Q_m$ increases, showing again that the magnetic charge introduces measurable dynamical effects. 

In addition to the geometrical and dynamical analysis, we confronted the model with observational twin--peak QPO data from four sources covering a wide range of mass scales: the stellar--mass systems XTE~J1550--564 and GRO~J1655--40, the intermediate--mass candidate M82~X--1, and the supermassive black hole Sgr~A$^\ast$. By performing a two--dimensional $\Delta\chi^2$ analysis in the $(r,Q_m)$ parameter space, we determined the best--fit values and their corresponding confidence regions. For all sources, the minimum of the likelihood occurs at $Q_m = 0$, indicating that the present QPO data do not require a nonvanishing magnetic charge. However, finite values of $Q_m$ remain allowed within the statistical uncertainties. From the $1\sigma$ contours, we obtain an upper bound $Q_m/M \lesssim 0.2$ within the explored domain, while the preferred orbital radius $r$ differs among the sources in the range $r/M \sim 8$--$12$. 

Therefore, although the magnetic charge produces clear theoretical modifications in photon motion and particle dynamics, the current QPO observations primarily constrain the orbital radius and provide only moderate bounds on $Q_m$. Future high--precision timing observations may improve these constraints and help to further test the possible existence of magnetic charge in astrophysical black holes.

\section*{Acknowledgments}

F.A. acknowledges the Inter University Center for Astronomy and Astrophysics (IUCAA), Pune, India for granting visiting associateship. 

\bibliographystyle{apsrev4-2}
\bibliography{references}

\end{document}